\documentclass[journal,10pt,twocolumn]{IEEEtran}

\pdfoutput=1

\usepackage[cmex10]{amsmath}
\usepackage{amssymb,amsthm,bm}  
\usepackage{calc}  
\usepackage{cite}  
\usepackage{color}  
\usepackage{graphicx}  
\usepackage{ifthen}  
\usepackage{ifpdf}  
\usepackage[update,suffix=-converted]{epstopdf}  
\usepackage{mathtools}  

\theoremstyle{plain}
\newtheorem{theorem}{Theorem}
\newtheorem{corollary}{Corollary}

\newtheorem{definition}{Definition}

\renewcommand{\figurename}{\footnotesize Figure}
\renewcommand{\tablename}{\it Table}

\mathtoolsset{showonlyrefs,showmanualtags}
\usetagform{default}

\newcommand{\chref}[1]{Chapter~\ref{#1}}
\newcommand{\secref}[1]{Section~\ref{#1}}

\newcommand{\figref}[1]{Figure~\ref{#1}}
\newcommand{\tabref}[1]{Table~\ref{#1}}

\newcommand{\thmref}[1]{Theorem~\ref{#1}}

\newcommand{\term}[1]{{\bfseries#1}}  

\newcommand{\up}[1]{{\textup{#1}}}  
\newcommand{\dele}[1]{}  
\newcommand{\change}[2]{#2}  

\newcommand{\st}{:}  
\newcommand{\inv}[1]{{#1}^{-1}}  

\newcommand{\cross}{\times}  

\newcommand{\ppprime}[1]{\prime\prime\prime}  

\newcommand{\pars}[1]{\left(#1\right)}  

\newcommand{\parsi}[1]{(#1)}  
\newcommand{\parsii}[1]{\bigl(#1\bigr)}  

\newcommand{\bracks}[1]{\left[#1\right]}  

\newcommand{\bracksi}[1]{[#1]}  
\newcommand{\bracksii}[1]{\bigl[#1\bigr]}  

\newcommand{\braces}[1]{\left\{#1\right\}}  

\newcommand{\reals}{\mathbb{R}}  
\newcommand{\complex}{\mathbb{C}}  

\newcommand{\abs}[1]{\left|#1\right|}  

\newcommand{\argmin}{\operatornamewithlimits{arg\,min}}

\newcommand{\setdiff}{\backslash}  

\newcommand{\vect}[1]{\boldsymbol{\mathrm{#1}}}  
\newcommand{\matr}[1]{\boldsymbol{\mathrm{#1}}}  
\newcommand{\matrels}[1]{\begin{bmatrix}#1\end{bmatrix}}  
\newcommand{\idmat}{\matr{I}}  

\newcommand{\tp}[1]{#1^{T}}  
\newcommand{\pinv}[1]{#1^{+}}  
\newcommand{\herm}[1]{#1^{H\!}}  

\newcommand{\tr}{\textup{\textrm{tr}}}  

\newcommand{\norm}[1]{\left\|#1\right\|}  

\newcommand{\normii}[1]{\bigl\|#1\bigr\|}  

\newcommand{\rank}{\textup{\textrm{rank}}}  

\newcommand{\pdef}{\succeq}  
\newcommand{\PDEF}{\succ}  

\newcommand{\schur}[1]{\overline{#1}}  

\newcommand{\ev}{\mathbb{E}}  

\newcommand{\distas}{\sim}  

\newcommand{\given}{\left.\right|}  

\newcommand{\givenii}{\,\bigl.\bigr|\,}  
\newcommand{\giveniii}{\,\Bigl.\Bigr|\,}  

\newcommand{\normpdf}{\mathcal{N}}  

\newcommand{\bigO}{O}  

\newcommand{\allin}{:}  

\newlength{\templength}

\newcommand{\Nystrom}{Nystr\"{o}m}

\newcommand{\NystromCovEst}{\widehat{\matr{\Sigma}}\parsi{I}}

\newcommand{\MSE}{\textup{\textrm{MSE}}}

\renewcommand{\IEEEQED}{\IEEEQEDopen}
\renewcommand{\term}[1]{\emph{#1}}  

\begin{document}

	
\title{Estimating Principal Components of Covariance Matrices Using the \Nystrom\ Method}
\author{Nicholas~Arcolano~and~Patrick~J.~Wolfe%
\thanks{This work is sponsored by the United States Air Force under contract \mbox{FA8721-05-C-0002}. Opinions, interpretations, recommendations, and conclusions are those of the authors and are not necessarily endorsed by the United States Government.}}
\maketitle


\begin{abstract}
		Covariance matrix estimates are an essential part of many signal processing algorithms, and are often used to determine a low-dimensional principal subspace via their spectral decomposition.  However, exact eigenanalysis is computationally intractable for sufficiently high-dimensional matrices, and in the case of small sample sizes, sample eigenvalues and eigenvectors are known to be poor estimators of their population counterparts.  To address these issues, we propose a covariance estimator that is computationally efficient while also performing shrinkage on the sample eigenvalues.  Our approach is based on the \Nystrom\ method, which uses a data-dependent orthogonal projection to obtain a fast low-rank approximation of a large positive semidefinite matrix.  We provide a theoretical analysis of the error properties of our estimator as well as empirical results, including examples of its application to adaptive beamforming and image denoising.
\end{abstract}


\section{Introduction} \label{sec:Intro}

\IEEEPARstart{T}{he} need to determine a principal subspace containing some signal of interest arises in many areas of signal processing, including beamforming \cite{Cardoso1993}, speech processing \cite{Ephraim1995}, and source separation \cite{Belouchrani1997}.  Typically, subspace estimation involves computing the spectral decomposition of a covariance matrix that has been estimated using a set of observed data vectors.  The estimator most commonly used in practice is the \term{sample covariance}, often preferred because it is simple to compute and has well-understood theoretical properties.

When solving the subspace estimation problem, one faces two critical challenges.  The first is that for $p$-dimensional data, the computational cost of the full spectral decomposition scales as $\bigO(p^3)$.  In the case of high-dimensional data sets, or when the subspace estimation problem needs to be solved many times, obtaining the eigenvalues and eigenvectors of the sample covariance becomes a computational bottleneck.  For this reason, algorithms have been developed to obtain approximate solutions to the eigenvalue problem \cite{Lanczos1950,Arnoldi1951}.

The second challenge is that the eigenvalues and eigenvectors the sample covariance matrix are known to be be poor estimates of the true eigenvalues and eigenvectors, especially when operating in high dimensions with limited observations \cite{Marcenko1967,Johnstone2001}.  In particular, the sample eigenvalues are known to be over-dispersed (relative to the true spectrum), and many researchers have focused on developing \term{shrinkage estimators} that yield improved estimation results \cite{Stein1975,Haff1980,Dey1985,Ledoit2004}.

Instead of addressing these challenges separately, we propose that by solving them concurrently, one can perform both tasks at a reduced computational cost.  To this end, we develop an estimator based on the \term{\Nystrom\ method} \cite{Williams2000,Fowlkes2004,Arcolano2011}, a matrix approximation technique that uses a data-dependent orthogonal projection to approximate a positive semidefinite matrix.  This approach leads to an estimator that not only admits computationally efficient spectral analysis, but also shrinks the eigenvalues of the sample covariance. 

We begin by formulating the covariance estimation problem and reviewing the \Nystrom\ method for matrix approximation. We then develop its use as a covariance estimator, including a study of its error characteristics.  We conclude with examples of the use the \Nystrom\ covariance estimator in two practical applications: adaptive beamforming and image denoising.
\IEEEpubidadjcol

\section{The Covariance Estimation Problem} \label{sec:CovarianceEstimation}

Let $\matr{X}$ be a $p \times n$ matrix whose columns $\vect{x}_1,\dotsc,\vect{x}_n$ are independent and identically distributed (i.i.d.) samples from an unknown $p$-variate distribution.  Throughout the following, we assume $\vect{x}_1,\dotsc,\vect{x}_n$ have zero mean and a finite covariance, denoted by the \mbox{$p \times p$} positive semidefinite matrix $\matr{\Sigma}$.\footnote{We refer to a $p \times p$ matrix $\matr{\Sigma}$ as \term{positive semidefinite} (denoted $\matr{\Sigma} \pdef 0$) if it is symmetric and $\tp{\vect{x}}\matr{\Sigma}\vect{x} \geq 0$ for all $\vect{x} \in \reals^n$, and as \term{positive definite} (denoted $\matr{\Sigma} \PDEF 0$) if it is positive semidefinite with $\tp{\vect{x}}\matr{\Sigma}\vect{x} = 0$ if and only if $\vect{x} = \vect{0}$.}

The basic problem of covariance estimation is straightforward: given $\matr{X}$, we wish to construct an estimator $\widehat{\matr{\Sigma}}$ of $\matr{\Sigma}$.  As a function of random data, $\widehat{\matr{\Sigma}}$ is itself random, and thus its performance as an estimator is best understood through its statistical properties (conditional on the true covariance).  In particular, we will be concerned with the \term{bias matrix}
\begin{equation}
	\matr{B}\parsii{\widehat{\matr{\Sigma}} \givenii \matr{\Sigma}}
		\equiv \matr{\Sigma} - \ev\parsii{\widehat{\matr{\Sigma}} \givenii \matr{\Sigma}},
\end{equation}
and the \term{mean squared error} (MSE)
\begin{equation} \label{eq:MeanSquaredErrorDef}
	\MSE\parsii{\widehat{\matr{\Sigma}} \givenii \matr{\Sigma}}
		\equiv \ev\pars{\normii{\matr{\Sigma} - \widehat{\matr{\Sigma}}}^2 \giveniii \matr{\Sigma}},
\end{equation}
where $\norm{\cdot}$ is a suitable matrix norm.  A common choice of norm is the Frobenius norm, defined for a real matrix $\matr{A}$ as $\norm{\matr{A}}_F = \bracksi{\tr\parsi{\tp{\matr{A}}\matr{A}}}^{1/2}$.  This norm is used throughout the covariance estimation literature \cite{Leung1987,Ledoit2004}, and will be the primary one featured in our analysis. 

\dele{Often a ``good'' estimator is identified with having minimal bias and a low MSE.  However, depending on the application, we may wish to balance raw error performance with other desirable qualities, such as computational efficiency, sparsity, invertibility, or favorable spectral characteristics.  We will primarily be concerned with estimators that possess two important properties: first, their eigenvalues and eigenvectors can be computed efficiently, and second, their eigenvalues are effective estimators of the true eigenvalues of $\matr{\Sigma}$.}

\subsection{The Sample Covariance}

The most common covariance estimator is the \term{sample covariance matrix},
\begin{equation}
	\matr{S} = \frac{1}{n} \sum_{i=1}^n \vect{x}_i \tp{\vect{x}}_i
		= \frac{1}{n} \matr{X} \tp{\matr{X}}.
\end{equation}
This estimator has a number of qualities that make it a popular choice among practitioners.  For example, it is unbiased, and its computational cost of $\bigO(p^2 n)$ is not excessively expensive.  When $\vect{x}_1,\dotsc,\vect{x}_n$ are i.i.d.\ samples from the $p$-variate normal distribution with mean $\vect{0}$ and positive definite covariance $\matr{\Sigma}$---denoted $\normpdf_p(\vect{0},\matr{\Sigma})$---it corresponds to the maximum-likelihood estimator of $\matr{\Sigma}$ given the data.  We can also compute its MSE with respect to the Frobenius norm, given by
\begin{align} \label{eq:SampleCovarianceMSE}
	\MSE\pars{\matr{S} \given \matr{\Sigma}}
		\equiv \frac{1}{n} \bracks{\tr\parsii{\matr{\Sigma}^2}
		+ \tr^2\parsii{\matr{\Sigma}}}.
\end{align}

Despite these convenient properties, there are characteristics of $\matr{S}$ that make it unsuitable for many applications.  For example, if $n < p$, then $\matr{S}$ is guaranteed to be rank-deficient; even if we have prior reason to believe that $\matr{\Sigma}$ is invertible, its estimate will be singular.  Another issue is the over-dispersion of the sample eigenvalues.  Let $\lambda_1(\matr{S}),\dotsc,\lambda_p(\matr{S})$ denote the eigenvalues of $\matr{S}$ in nonincreasing order.  For fixed $p$ and large $n$, the sample eigenvalues are reasonable estimators of the true eigenvalues of $\matr{\Sigma}$, and it can be shown \cite{Anderson1963} that as $n \to \infty$ with $p$ fixed, $\lambda_i(\matr{S})$ converges almost surely to $\lambda_i(\matr{\Sigma})$ for $i = 1,\dotsc,p$.  However, when $p$ is allowed to grow with $n$ (keeping the ratio $n/p$ fixed), results such as the celebrated \emph{Mar\v{c}kenko-Pastur law} suggest that the sample eigenvalues are not effective estimators, and fail to converge to the true eigenvalues \cite{Marcenko1967,Johnstone2001}.  

\subsection{Shrinkage Covariance Estimators}

In the context of covariance estimation, \term{shrinkage estimation} involves compensating for the known over-dispersion of sample eigenvalues, in order to improve error performance and numerical stability.  One common approach to covariance shrinkage is to preserve the sample eigenvectors, but alter the sample eigenvalues to improve error performance with respect to a given loss function \cite{Stein1975,Dey1985}.  \dele{
For example, in \cite{Dey1985}, the authors derive a minimax-optimal estimator that shrinks the sample eigenvalues as
\begin{equation} \label{eq:MinimaxCovarianceEstimator}
	\lambda_i\parsii{\widehat{\matr{\Sigma}}} = \frac{n}{n + p + 1 - 2i}\,
		\lambda_i\parsii{\matr{S}}, \quad i = 1,\dotsc,p.
\end{equation}}
Although shrinkage estimators of this form often come with analytical guarantees regarding error performance, they do so at the cost of operating on the spectral decomposition of $\matr{S}$ directly.  Consequently, such estimators are impractical for large data sets.

Another approach is to construct the estimate $\widehat{\matr{\Sigma}}$ as a linear combination of $\matr{S}$ and a known positive definite matrix \cite{Haff1980,Ledoit2004}.  For example, the Ledoit-Wolf estimator of \cite{Ledoit2004} takes the form
\begin{equation} \label{eq:LedoitWolfCovarianceEstimator}
	\widehat{\matr{\Sigma}} = \alpha \matr{S} + \beta \idmat_p,
\end{equation}
where the optimal shrinkage coefficients $\alpha,\beta \geq 0$ are given by
\begin{equation}
	\pars{\alpha^{\ast}, \beta^{\ast}} = \argmin_{(\alpha,\beta)}
		\norm{\pars{\alpha \matr{S} + \beta \idmat_p} - \matr{\Sigma}}_F^2.
\end{equation}
Since the optimal coefficients can only be obtained analytically in the case where $\matr{\Sigma}$ is known, the authors instead develop consistent estimators for $\alpha$ and $\beta$ based on the data.  Note that while estimators such as \eqref{eq:LedoitWolfCovarianceEstimator} have lower computational demands than those that operate on the eigenvalues of $\matr{S}$ directly, once we have obtained an estimate $\widehat{\matr{\Sigma}}$, we still must pay the full cost of $\bigO(p^3)$ if we wish to obtain its principal components.

\subsection{Covariance Estimation Using the \Nystrom\ Method}

Let $I \subseteq \{1,\dotsc,p\}$ be a set of $k$ indices.  The \Nystrom\ covariance estimator \cite{Arcolano2011} takes the form
\begin{equation} \label{eq:NystromCovarianceEstimatorIntro}
	\widehat{\matr{\Sigma}}\pars{I} = \frac{1}{n}\matr{X}\,\matr{P}\!\pars{I}\tp{\matr{X}},
\end{equation}
where $\matr{P}\pars{I}$ represents an orthogonal projection onto the subspace of $\reals^n$ spanned by the $k$ rows of $\matr{X}$ specified by the indices in $I$.  

Defining $\widehat{\matr{\Sigma}}$ in this fashion serves two important purposes.  First, we will show that \eqref{eq:NystromCovarianceEstimatorIntro} is equivalent to the \term{\Nystrom\ approximation} of the sample covariance, an established method for low-rank approximation of positive semidefinite matrices \cite{Williams2000,Fowlkes2004}.  A primary advantage of this method is its computational efficiency, as one can obtain the $k$ principal eigenvalues and eigenvectors of $\widehat{\matr{\Sigma}}$ for a cost that scales linearly in $p$ and $n$.  Second, the projection $\matr{P}$ shrinks the singular values of the data, serving to counteract over-dispersion of eigenvalues in the sample covariance.

\dele{We continue with a brief review of the \Nystrom\ method.  We then develop the \Nystrom\ covariance estimator further, deriving expressions for its bias and MSE, and exploring its spectral characteristics and computational properties.}

\section{The \Nystrom\ Method} \label{sec:Nystrom}

The \Nystrom\ method is a classical technique for obtaining numerical solutions to eigenfunction problems.  When applied to matrices, it can be used to construct a low-rank approximation of a positive semidefinite matrix as follows.

Let $\matr{Q}$ be a $p \times p$ positive semidefinite matrix, represented in block form as
\begin{equation} 
	\matr{Q} = \matrels{\matr{Q}_{11} & \matr{Q}_{12} \\ \tp{\matr{Q}}_{12} & \matr{Q}_{22}},
\end{equation}
where $\matr{Q}_{11}$ is $k \times k$.  The \Nystrom\ approximation of $\matr{Q}$ preserves $\matr{Q}_{11}$ and $\matr{Q}_{12}$ while approximating $\matr{Q}_{22}$ by its \term{\Nystrom\ extension}:
\begin{equation} \label{eq:NystromApproximation}
	\widehat{\matr{Q}} \equiv \matrels{\matr{Q}_{11} & \matr{Q}_{12} \\ \tp{\matr{Q}}_{12} & \tp{\matr{Q}}_{12}\pinv{\matr{Q}}_{11}\matr{Q}_{12}},
\end{equation}
where $\pinv{\matr{Q}}_{11}$ denotes the Moore-Penrose pseudoinverse of $\matr{Q}_{11}$.  Since the approximation reconstructs $\matr{Q}_{11}$ and $\matr{Q}_{12}$ exactly, the approximation error $\matr{Q} - \widehat{\matr{Q}}$ is characterized entirely by the \term{Schur complement} of $\matr{Q}_{11}$ in $\matr{Q}$, 
\begin{equation}  
	\schur{\matr{Q}}_{11} \equiv \matr{Q}_{22} - \tp{\matr{Q}}_{12}\pinv{\matr{Q}}_{11}\matr{Q}_{12}.
\end{equation}

If we view $\matr{Q}$ as the outer product of an underlying data matrix, an alternative way to characterize the \Nystrom\ approximation is as a function of an orthogonal projection.  Let $\matr{X}$ be a $p \times n$ matrix, partitioned as
\begin{equation} \label{eq:XPartition}
	\matr{X} = \matrels{\matr{Y} \\ \matr{Z}} 
\end{equation}
where $\matr{Y}$ is $k \times n$ and $\matr{Z}$ is $(p-k) \times n$, and let 
\begin{equation} 
	\matr{Q} = \matr{X}\tp{\matr{X}}
		= \matrels{\matr{Y}\tp{\matr{Y}} & \matr{Y}\tp{\matr{Z}} \\
		\matr{Z}\tp{\matr{Y}} & \matr{Z}\tp{\matr{Z}}}.
\end{equation}
We then define the $n \times n$ symmetric idempotent matrix
\begin{equation} \label{eq:NystromProjection}
	\matr{P} \equiv \tp{\matr{Y}}\pinv{\parsii{\tp{\matr{Y}}}}
		= \tp{\matr{Y}}\pinv{\parsii{\matr{Y}\tp{\matr{Y}}}}\matr{Y},
\end{equation}
which represents an orthogonal projection onto the subspace of $\reals^n$ spanned by the $k$ rows of $\matr{Y}$.  We obtain the same expression as in \eqref{eq:NystromApproximation} by approximating $\matr{X}$ with its projection $\matr{X}\matr{P}$:
\begin{align} 
	\widehat{\matr{Q}} = \matr{X}\matr{P}\tp{\parsi{\matr{X}\matr{P}}}
	& = \matr{X}\matr{P}\tp{\matr{X}} \\
	& = \matrels{\matr{Y}\tp{\matr{Y}} & \matr{Y}\tp{\matr{Z}} \\
		\matr{Z}\tp{\matr{Y}} & 
		\matr{Z}\tp{\matr{Y}}\pinv{\parsii{\matr{Y}\tp{\matr{Y}}}}\matr{Y}\tp{\matr{Z}}}.
\end{align}
This interpretation illustrates the low-rank nature of $\widehat{\matr{Q}}$, as we must have $\rank\parsii{\widehat{\matr{Q}}} \leq \rank\pars{\matr{P}} \leq k$.  It also highlights the fact that the approximation need not be restricted to the first $k$ rows of $\matr{X}$; we may instead choose to construct $\matr{P}$ based on \emph{any} subset of $k$ rows.  Since different subsets typically yield different approximations, we can view $\widehat{\matr{Q}}$ as a function of a set of $k$ indices $I \subseteq \{1,\dotsc,p\}$.  (Throughout the article, we we will use $\matr{A}_{IJ}$ to denote the submatrix of a matrix $\matr{A}$ whose rows and columns are specified by respective index sets $I$ and $J$, and define $\matr{A}_{I} \equiv \matr{A}_{II}$.)

The problem of selecting a suitable subset for \Nystrom\ approximation is one that has received significant attention in the literature.  Although a number of efforts have focused on developing advanced subset selection methods \cite{Drineas2005,Belabbas2009a}, for many applications satisfactory performance can be achieved simply by choosing $I$ randomly with uniform probability \cite{Williams2000,Fowlkes2004}.  This simpler approach has the added benefit of enhancing the computational gains associated with the \Nystrom\ method, and will be the strategy employed for the experiments in Sections~\ref{sec:Beamforming} and \ref{sec:Denoising}.

\section{The \Nystrom\ Covariance Estimator} \label{sec:NystromCov}

We proceed by formally defining the \Nystrom\ covariance estimator, after which we derive expressions for its bias and MSE.  We then discuss its eigenvalue shrinkage properties and derive expressions for its eigenvalues and eigenvectors.


\begin{definition}[\Nystrom\ covariance estimator] \label{def:NystromCovarianceEstimator}
	Let $\matr{X}$ be a $p \times n$ matrix whose columns $\vect{x}_1,\dotsc,\vect{x}_n$ are i.i.d.\ random vectors such that $\ev\parsi{\vect{x}_i} = \vect{0}$ and $\ev\parsi{\vect{x}_i\tp{\vect{x}}_i} = \matr{\Sigma}$ for $i = 1,\dotsc,n$.  Let $\vect{r}_1,\dotsc,\vect{r}_p$ denote the rows of $\matr{X}$.  Given a $k$-subset \mbox{$I \subseteq \{1,\dotsc,p\}$}, we define the \Nystrom\ covariance estimator of $\matr{\Sigma}$ as 
\begin{equation} 
	\NystromCovEst \equiv {\frac{1}{n}}\matr{X}\,\matr{P}\!\pars{I}\tp{\matr{X}},
\end{equation}
where $\matr{P}\pars{I}$ represents an orthogonal projection onto the subspace of $\reals^n$ spanned by the set of vectors $\braces{\vect{r}_i \st i \in I}$.
\end{definition}


As previously discussed, $\NystromCovEst$ is a function of an index set $I \subseteq \{1,\dotsc,p\}$, and thus error performance will conditional on $I$.  Although viewed here as an estimator of $\matr{\Sigma}$, the \Nystrom\ covariance estimator could be interpreted as the \Nystrom\ approximation of the sample covariance $\matr{S}$.  When $\rank\!\pars{\matr{P}\pars{I}} = \rank\!\pars{\matr{X}} \leq \min(p,n)$, this approximation is exact, and we have $\NystromCovEst = \matr{S}$.

\subsection{Error Statistics}

Assume now that the columns of $\matr{X}$ are drawn independently from a \mbox{$p$-variate} normal distribution with zero mean and covariance $\matr{\Sigma} \PDEF 0$.  In this case, we can derive analytical expressions for the bias and expected square error of the \Nystrom\ covariance estimator.  We begin by computing the expected value of $\NystromCovEst$, after which the bias matrix follows as a corollary.


\begin{theorem}[Expected value of \Nystrom\ covariance estimator]
\label{thm:NystromCovarianceEV}
	Let $\matr{X}$ be a $p \times n$ matrix whose columns $\vect{x}_1,\dotsc,\vect{x}_n$ are i.i.d.\ random samples from $\normpdf_p\parsi{\vect{0},\matr{\Sigma}}$.  Let $\NystromCovEst$ be the \Nystrom\ covariance estimator of $\matr{\Sigma}$ given a $k$-subset \mbox{$I \subseteq \{1,\dotsc,p\}$}, and define \mbox{$J = \{1,\dotsc,p\} \setminus I$}.  Then,
$\bracksii{\ev\parsii{\NystromCovEst}}_I = \bracks{\matr{\Sigma}}_I$, $\bracksii{\ev\parsii{\NystromCovEst}}_{IJ} = \tp{\bracksii{\ev\parsii{\NystromCovEst}}}_{JI} = \bracks{\matr{\Sigma}}_{IJ}$, and
\begin{equation}
	\bracks{\ev\parsii{\NystromCovEst}}_J = 
		\frac{k}{n}\matr{\Sigma}_{J}
			+ \frac{n-k}{n}\tp{\matr{\Sigma}}_{IJ}\inv{\matr{\Sigma}}_{J}\matr{\Sigma}_{IJ}\,.
\end{equation}
\end{theorem}


\begin{IEEEproof}
	Without loss of generality, let $I = \{1,\dotsc,k\}$ and $J = \{k+1,\dotsc,p\}$.  Partitioning $\matr{X}$ as in \eqref{eq:XPartition}, the \Nystrom\ covariance estimate is given by
\begin{equation} 
	\widehat{\matr{\Sigma}} = \frac{1}{n}\matr{X}\matr{P}\tp{\matr{X}}
		= \frac{1}{n}\matrels{\matr{Y}\tp{\matr{Y}} & \matr{Y}\tp{\matr{Z}} \\
		\matr{Z}\tp{\matr{Y}} & \matr{Z}\matr{P}\tp{\matr{Z}}},
\end{equation}
where $\matr{P}$ represents an orthogonal projection onto the span of the rows of $\matr{Y}$.  By construction, we have
\begin{align}
	\ev\parsii{{\textstyle\frac{1}{n}}\matr{Y}\tp{\matr{Y}}} & = \matr{\Sigma}_I\,,
\end{align}
and
\begin{align}
	\ev\parsii{{\textstyle\frac{1}{n}}\matr{Y}\tp{\matr{Z}}} &
		 = \tp{\bracksii{\ev\parsii{{\textstyle\frac{1}{n}}\matr{Z}\tp{\matr{Y}}}}}
		 = \matr{\Sigma}_{IJ}\,,
\end{align}
and thus we need only compute $\ev\parsii{{\textstyle\frac{1}{n}}\matr{Z}\matr{P}\tp{\matr{Z}}}$.  To perform this calculation, consider the nested expectation
\begin{equation} 
	\ev\parsii{\matr{Z}\matr{P}\tp{\matr{Z}}}
		= \ev_Y\!\bracks{\ev\parsii{\matr{Z}\matr{P}\tp{\matr{Z}}\!\given \matr{Y}}}.
\end{equation}
Using standard properties of conditional distributions of normal random vectors, one can show that given $\matr{Y}$, the columns $\vect{z}_1,\dotsc,\vect{z}_n$ of $\matr{Z}$ are independent and normally distributed as
\begin{equation}
	\vect{z}_i \given \matr{Y} \distas \normpdf_{p-k}\pars{\vect{\mu}_{Z|Y}^{(i)},\matr{\Sigma}_{Z|Y}},
\end{equation}
where
\begin{equation}
	\vect{\mu}_{Z|Y}^{(i)} = \tp{\matr{\Sigma}}_{IJ}\inv{\matr{\Sigma}}_{I} \matr{y}_i\,,
\end{equation}
and
\begin{equation}
	\matr{\Sigma}_{Z|Y}
		= \matr{\Sigma}_{J} - \tp{\matr{\Sigma}}_{IJ}\inv{\matr{\Sigma}}_{I}\matr{\Sigma}_{IJ}
		= \schur{\matr{\Sigma}}_I.
\end{equation}
Given these distributions, evaluating $\ev\parsii{\matr{Z}\matr{P}\tp{\matr{Z}}\!\given \matr{Y}}$ is a matter of performing standard moment calculations using the properties of normal random variables.  For convenience, we apply a result from \cite[Theorem~2.3.5]{Gupta2000} for normal random matrices, which states that for a random $p \times n$ matrix $\matr{X}$ whose columns $\vect{x}_1,\dotsc,\vect{x}_n$ are distributed as $\vect{x}_i \distas \normpdf_{p}\pars{\vect{\mu}_i,\matr{\Sigma}}$, if $\matr{A}$ is a constant $p \times p$ matrix, then
\begin{equation}
	\ev\parsii{\matr{X}\matr{A}\tp{\matr{X}}}
		= \tr\pars{\matr{A}}\matr{\Sigma} + \matr{M}\matr{A}\tp{\matr{M}},
\end{equation}
where $\matr{M} = [\vect{\mu}_1\ \cdots\ \vect{\mu}_n]$.  Thus, 
\begin{equation}
	\ev\parsii{\matr{Z}\matr{P}\tp{\matr{Z}}\!\given \matr{Y}}
		= k\schur{\matr{\Sigma}}_I
		+ \tp{\matr{\Sigma}}_{IJ}\inv{\matr{\Sigma}}_{I}\matr{Y}\matr{P}\tp{\matr{Y}}
		\inv{\matr{\Sigma}}_{I}\matr{\Sigma}_{IJ}\,,
\end{equation}
and
\begin{align}
	\ev\parsii{\matr{Z}\matr{P}\tp{\matr{Z}}}
	& = \ev_Y\!\bracks{\ev\parsii{\matr{Z}\matr{P}\tp{\matr{Z}}\!\given \matr{Y}}} \\ 
	& = \ev_Y\!\bracks{k\schur{\matr{\Sigma}}_I
		+ \tp{\matr{\Sigma}}_{IJ}\inv{\matr{\Sigma}}_{I}\matr{Y}\matr{P}\tp{\matr{Y}}
		\inv{\matr{\Sigma}}_{I}\matr{\Sigma}_{IJ}} \\
	& = k\matr{\Sigma}_J + \pars{n-k}\tp{\matr{\Sigma}}_{IJ}\inv{\matr{\Sigma}}_{I}\matr{\Sigma}_{IJ},
\end{align}
where the final equality follows from $\ev_Y\parsii{\matr{Y}\matr{P}\tp{\matr{Y}}} = n \matr{\Sigma}_I$.  Dividing by $n$ yields the desired result.
\end{IEEEproof}


\begin{corollary}[Bias of \Nystrom\ covariance estimator \cite{Arcolano2011}]
\label{cor:NystromCovarianceBias}
	Let $\matr{X}$ be a $p \times n$ matrix whose columns $\vect{x}_1,\dotsc,\vect{x}_n$ are i.i.d.\ random samples from $\normpdf_p\parsi{\vect{0},\matr{\Sigma}}$.  Let $\NystromCovEst$ be the \Nystrom\ covariance estimator of $\matr{\Sigma}$ given a $k$-subset \mbox{$I \subseteq \{1,\dotsc,p\}$}, and define \mbox{$J = \{1,\dotsc,p\} \setminus I$}.  Then the bias matrix
\begin{equation}
	\matr{B}\parsii{\NystromCovEst \givenii \matr{\Sigma}}
		= \matr{\Sigma} - \ev\parsii{\NystromCovEst}
\end{equation}
satisfies $[\matr{B}]_{ij} = 0$ for all $(i,j) \notin J \cross J$, and
\begin{equation}
	\matr{B}_J = \frac{n-k}{n}\schur{\matr{\Sigma}}_I 
		= \frac{n-k}{n}\bracks{\matr{\Sigma}_{J}
		- \tp{\matr{\Sigma}}_{IJ}\inv{\matr{\Sigma}}_{J}\matr{\Sigma}_{IJ}}.
\end{equation}
\end{corollary}


Thus, $\NystromCovEst$ is a biased estimator of $\matr{\Sigma}$, except in the case where the Schur complement $\schur{\matr{\Sigma}}_I = \matr{0}$.  Recalling from \secref{sec:Nystrom} that this Schur complement also expresses the error between $\matr{\Sigma}$ and its \Nystrom\ approximation, we see that $\NystromCovEst$ cannot be unbiased unless it is equal to the sample covariance.


\begin{theorem}[MSE of \Nystrom\ covariance estimator \cite{Arcolano2011}]
\label{thm:NystromCovarianceMSE}
	Let $\matr{X}$ be a $p \times n$ matrix whose columns $\vect{x}_1,\dotsc,\vect{x}_n$ are i.i.d.\ random samples from $\normpdf_p\parsi{\vect{0},\matr{\Sigma}}$.  Let $\NystromCovEst$ be the \Nystrom\ covariance estimator of $\matr{\Sigma}$ given a $k$-subset \mbox{$I \subseteq \{1,\dotsc,p\}$}, and define \mbox{$J = \{1,\dotsc,p\} \setminus I$}.  Then the mean square error of the \Nystrom\ covariance estimator in Frobenius norm is
\begin{align} 
	& \ev\,\normii{\matr{\Sigma} - \NystromCovEst}^2_F \\
	& \quad = \MSE\pars{\matr{S} \given \matr{\Sigma}}
		+ \frac{n-k}{n^2}
		\bracks{(n-k-1)\tr\parsii{\schur{\matr{\Sigma}}_I^2}
		- \tr^2\parsii{\schur{\matr{\Sigma}}_I}} \\[-2ex]
	& \quad = \MSE\pars{\matr{S} \given \matr{\Sigma}}
		+ \frac{(n-k)^2}{n^2}\bracks{\norm{\schur{\matr{\Sigma}}_I}_F^2
		- \MSE\pars{\schur{\matr{S}}_I \given \matr{\Sigma}}}, \\[-4ex]
\end{align}
where $\MSE\pars{\matr{S} \given \matr{\Sigma}}$ is the mean square error of the sample covariance estimator \change{$\matr{S} = \frac{1}{n}\matr{X}\tp{\matr{X}}$, given by
\begin{equation} 
	\MSE\pars{\matr{S} \given \matr{\Sigma}}
		\equiv \frac{1}{n} \bracks{\tr\parsii{\matr{\Sigma}^2}
		+ \tr^2\parsii{\matr{\Sigma}}},
\end{equation}}
{given in \eqref{eq:SampleCovarianceMSE},} and $\MSE\pars{\schur{\matr{S}}_I \given \matr{\Sigma}}$ is the mean square error of the sample covariance estimator of the Schur complement of $\matr{\Sigma}_I$ in $\matr{\Sigma}$, given by
\begin{equation} 
	\MSE\pars{\schur{\matr{S}}_I \given \matr{\Sigma}}
		\equiv \frac{1}{n-k} \bracks{\tr\parsii{\schur{\matr{\Sigma}}_I^2}
		+ \tr^2\parsii{\schur{\matr{\Sigma}}_I}}.
\end{equation}
\end{theorem}


\medskip
\begin{IEEEproof}
	Let \mbox{$I = \{1,\dotsc,k\}$} and \mbox{$J = \{k+1,\dotsc,p\}$} without loss of generality.  The MSE in Frobenius norm is
\begin{align}
	\ev\,\normii{\matr{\Sigma} - \NystromCovEst}^2_{F}
	& = \tr\parsii{\matr{\Sigma}^2}
		- 2\,\tr\!\bracks{\matr{\Sigma}\,
		\ev\pars{{\textstyle\frac{1}{n}}\matr{X}\matr{P}\tp{\matr{X}}}} \\
 	\label{eq:NystromCovarianceFrobeniusError}
 	& \qquad + \tr\!\bracks{\ev\pars{{\textstyle\frac{1}{n^2}}\matr{X}\matr{P}
		\tp{\matr{X}}\matr{X}\matr{P}\tp{\matr{X}}}}. 
\end{align}
Substituting the expression for $\ev\pars{{\frac{1}{n}}\matr{X}\matr{P}\tp{\matr{X}}}$ from \thmref{thm:NystromCovarianceEV}, we have
\begin{align}
	\tr\!\bracks{\matr{\Sigma}\,
		\ev\pars{{\displaystyle\frac{1}{n}}\matr{X}\matr{P}\tp{\matr{X}}}}
	& = \tr\!\pars{\matr{\Sigma}_I^2} + 2\,\tr\!\pars{\matr{\Sigma}_{IJ}\tp{\matr{\Sigma}}_{IJ}}
		+ \textstyle{\frac{k}{n}\tr\!\pars{\matr{\Sigma}_{J}^2}} \\[-2ex]
 	\label{eq:NystromCovarianceErrorPart1}
 	& \quad + {\textstyle\frac{(n-k)}{n}}
		\tr\!\pars{\matr{\Sigma}_{J}\tp{\matr{\Sigma}}_{IJ}
		\inv{\matr{\Sigma}}_{I}\matr{\Sigma}_{IJ}}.
\end{align}
To compute $\ev\pars{{\textstyle\frac{1}{n^2}}\matr{X}\matr{P}\tp{\matr{X}}\matr{X}\matr{P}\tp{\matr{X}}}$, let $\matr{X}$ be partitioned as in \eqref{eq:XPartition}, so that
\begin{align}
	\tr\parsii{\matr{X}\matr{P}\tp{\matr{X}}\matr{X}\matr{P}\tp{\matr{X}}}
	& = \tr\,\parsii{\matr{Y}\tp{\matr{Y}}\matr{Y}\tp{\matr{Y}}}
		+ 2\,\tr\parsii{\matr{Y}\tp{\matr{Z}}\matr{Z}\tp{\matr{Y}}} \\
	 \label{eq:MSEProofStep1}
	 & \quad + \tr\parsii{\matr{Z}\matr{P}\tp{\matr{Z}}\matr{Z}\matr{P}\tp{\matr{Z}}}.
\end{align}
As in the proof of \thmref{thm:NystromCovarianceEV}, the expectation of each term can be evaluated using standard properties of normal random vectors.  However, we may simplify the analysis using a result from \cite[Theorem~2.3.8]{Gupta2000}, which states that for a random $p \times n$ matrix $\matr{X}$ whose columns $\vect{x}_1,\dotsc,\vect{x}_n$ are distributed as $\vect{x}_i \distas \normpdf_{p}\pars{\vect{\mu}_i,\matr{\Sigma}}$, if $\matr{A}$, $\matr{B}$, and $\matr{C}$ are independent $n \times n$, $p \times p$, and $n \times n$ matrices (respectively), then
\begin{align}
	& \ev\pars{\matr{X}\matr{A}\tp{\matr{X}}\matr{B}\matr{X}\matr{C}\tp{\matr{X}}} \\
	& \quad = \tr\parsii{\tp{\matr{C}}\tp{\matr{A}}} \tr\pars{\matr{B}\matr{\Sigma}}\matr{\Sigma}
		+ \tr\pars{\matr{A}}\tr\pars{\matr{C}} \matr{\Sigma}\matr{B}\matr{\Sigma} \\
	& \qquad + \tr\parsii{\matr{A}\tp{\matr{C}}} \matr{\Sigma}\tp{\matr{B}}\matr{\Sigma}
		+ \tr\pars{\matr{C}}\matr{M}\matr{A}\tp{\matr{M}}\matr{B}\matr{\Sigma} \\
	& \qquad + \matr{M}\matr{A}\tp{\matr{C}}\tp{\matr{M}}\tp{\matr{B}}\matr{\Sigma}
		+ \tr\parsii{\matr{A}\tp{\matr{M}}\matr{B}\matr{M}\matr{C}}\matr{\Sigma} \\
	& \qquad + \tr\pars{\matr{B}\matr{\Sigma}} \matr{M}\matr{A}\matr{C}\tp{\matr{M}}
		+ \matr{\Sigma}\tp{\matr{B}}\matr{M}\tp{\matr{A}}\matr{C}\tp{\matr{M}} \\
	& \qquad + \tr\pars{\matr{A}} \matr{\Sigma}\matr{B}\matr{M}\matr{C}\tp{\matr{M}}
		+ \matr{M}\matr{A}\tp{\matr{M}}\matr{B}\matr{M}\matr{C}\tp{\matr{M}},
\end{align}
where $\matr{M} = [\vect{\mu}_1\ \cdots\ \vect{\mu}_n]$.  We can compute the first term in \eqref{eq:MSEProofStep1} by applying this formula directly; for the remaining two terms, we must use iterated expectation as we did in the proof of \thmref{thm:NystromCovarianceEV}.  Evaluating these expectations yields
\begin{equation} \label{eq:NystromCovarianceErrorPart2}	\ev\,\tr\,\parsii{\matr{Y}\tp{\matr{Y}}\matr{Y}\tp{\matr{Y}}}
	= \parsii{n^2 + n}\,\tr\parsii{\matr{\Sigma}_I^2}
	+ n\,\tr^2\parsii{\matr{\Sigma}_I},
\end{equation}
\begin{align}
	\ev\,\tr\,\parsii{\matr{Y}\tp{\matr{Z}}\matr{Z}\tp{\matr{Y}}}
	& = \parsii{n^2 + n}\,\tr\parsii{\matr{\Sigma}_{IJ}\tp{\matr{\Sigma}_{IJ}}} \\
 	\label{eq:NystromCovarianceErrorPart3}
 	& \quad + n\,\tr\parsii{\matr{\Sigma}_I}\tr\parsii{\matr{\Sigma}_J},
\end{align}
and
\begin{align} 
	\hspace{-2em}
	\ev\,\tr\parsii{\matr{Z}\matr{P}\tp{\matr{Z}}\matr{Z}\matr{P}\tp{\matr{Z}}}
	& = \parsii{k^2 + k}\,\tr\parsii{\schur{\matr{\Sigma}}_I^2}
		+ k\,\tr^2\parsii{\schur{\matr{\Sigma}}_I} \\
	& \quad + 2n\parsii{k+1}\,\tr\parsii{\schur{\matr{\Sigma}}_I\matr{R}} \\
	& \quad + 2n\parsii{k+1}\,\tr\parsii{\schur{\matr{\Sigma}}_I}\tr\parsi{\matr{R}} \\
	\label{eq:NystromCovarianceErrorPart4}
	& \quad + \parsii{n^2 + n}\,\tr\parsii{\matr{R}^2} + k\,\tr^2\parsii{\matr{R}},
\end{align}
where $\matr{R} \equiv \tp{\matr{\Sigma}}_{IJ}\inv{\matr{\Sigma}}_{I}\matr{\Sigma}_{IJ}$.  Substituting \eqref{eq:NystromCovarianceErrorPart1} and \eqref{eq:NystromCovarianceErrorPart2}--\eqref{eq:NystromCovarianceErrorPart4} into \eqref{eq:NystromCovarianceFrobeniusError} and simplifying terms yields the result.
\end{IEEEproof} \noeqref{eq:NystromCovarianceErrorPart3}


\subsection{Discussion of Error Results}

Given an arbitrary covariance $\matr{\Sigma}$, we can derive a lower bound for the MSE of the \Nystrom\ covariance estimator as follows.  First, note that when applied to the spectrum of a positive semidefinite matrix $\matr{A}$, the Cauchy-Schwarz inequality implies that $\tr^2\pars{\matr{A}} \leq \tr\parsii{\matr{A}^2}\,\rank\pars{\matr{A}}$.  For $\matr{A} = \schur{\matr{\Sigma}}_I$, substituting this inequality into the error expression in \thmref{thm:NystromCovarianceMSE} yields
\begin{align} 
	\ev\,\normii{\matr{\Sigma} - \NystromCovEst}^2_F
	& \geq \MSE\pars{\matr{S} \given \matr{\Sigma}}
		+ {\textstyle\frac{n-k}{n^2}}
		\pars{n-k-1} \tr\parsii{\schur{\matr{\Sigma}}_I^2} \\[-2ex]
	& \qquad - {\textstyle\frac{n-k}{n^2}}
		\tr\parsii{\schur{\matr{\Sigma}}_I^2}\,
		\rank\parsii{\schur{\matr{\Sigma}}_I^2} \\[2ex]
	& = \MSE\pars{\matr{S} \given \matr{\Sigma}}
		+ {\textstyle\frac{(n-k)(n-p-1)}{n^2}}
		\tr\parsii{\schur{\matr{\Sigma}}_I^2}\,,
\end{align}
where $\rank\parsii{\schur{\matr{\Sigma}}_I^2} = p-k$.  Thus, $n \leq p$ is a necessary condition for the MSE of the \Nystrom\ covariance estimator to be less than that of the sample covariance.  

We can show that equality is achieved for this bound when $\matr{\Sigma} = \idmat_p$.  In this case, $\schur{\matr{\Sigma}}_I = \idmat_{p-k}$ for all $k$-subsets $I \subseteq \{1,\dotsc,p\}$, and thus 
\begin{align} 
	\ev\,\normii{\matr{\Sigma} - \NystromCovEst}^2_F
	& = \MSE\pars{\matr{S} \given \matr{\Sigma}}
		+ {\textstyle\frac{n-k}{n^2}} (n-k-1) (p-k) \\[-2ex]
	& \qquad - {\textstyle\frac{n-k}{n^2}}(p-k)^2 \\[2ex]
	& = \MSE\pars{\matr{S} \given \matr{\Sigma}}
		+ {\textstyle\frac{(n-k)(p-k)(n-p-1)}{n^2}}\,,
\end{align}
where $\MSE\pars{\matr{S} \given \matr{\Sigma}} = (p^2 + p)/n$.  Since the second term in the above summation is negative if and only if $n - p - 1 < 0$, we see that when estimating the identity covariance, the \Nystrom\ estimator achieves better error performance than the sample covariance for any $n \leq p$.  

This behavior may seem surprising, especially when noting that $n \leq p$ implies that the sample covariance will be of rank $n$ with probability one, while the \Nystrom\ estimator will be of rank $k \leq n$.  However, eigenvalue results such as the Mar\v{c}enko-Pastur theorem indicate that in this regime the sample covariance will be over-dispersed.  Even though the \Nystrom\ estimator has fewer nonzero eigenvalues than does the sample covariance, the shrinkage it provides on these eigenvalues compensates for its lower rank, resulting in better overall error.

\dele{
In light of this discussion,  \Nystrom\ covariance estimator may find effective use within the ``large $p$, small $n$'' regime.  We will examine this hypothesis further through empirical comparisons in Sections~\ref{sec:Beamforming} and \ref{sec:Denoising}.}

\subsection{Eigenvalue Shrinkage}

The error performance of the \Nystrom\ covariance estimator derives from its ability to shrink the over-dispersed eigenvalues of the sample covariance.  One way to understand this property is as a consequence of the well-known eigenvalue inequalities of Weyl, a form of which \cite[Corollary~III.2.3]{Bhatia1997} states that for any $p \times p$ symmetric matrix $\matr{A}$ and $p \times p$ positive semidefinite matrix $\matr{B}$,
\begin{equation} \label{eq:WeylMonotonicityTheorem}
	\lambda_i\pars{\matr{A} + \matr{B}} \geq \lambda_i\pars{\matr{A}},
\end{equation}
for $i = 1,\dotsc, p$.

Let $\matr{S} \pdef 0$ be a $p \times p$ sample covariance matrix, and let $\NystromCovEst$ be the corresponding \Nystrom\ covariance estimator given a $k$-subset $I \subseteq \{1,\dotsc,p\}$.  As in previous discussions, we may let $I = \{1,\dotsc,k\}$ without loss of generality. By positive semidefiniteness of the Schur complement, the error matrix
\begin{equation} 
	\matr{E} \equiv \matr{S} - \NystromCovEst
		= \matrels{\matr{0} & \matr{0} \\ \matr{0} & \schur{\matr{S}}_I}
\end{equation}
is also positive semidefinite.  Thus, \eqref{eq:WeylMonotonicityTheorem} implies that
\begin{equation}
	\lambda_i\parsii{\NystromCovEst + \matr{E}}
		= \lambda_i\parsii{\matr{S}}
		\geq \lambda_i\parsii{\NystromCovEst},
\end{equation}
for $i = 1,\dotsc,p$.  In other words, given any $k$-subset, the \Nystrom\ covariance estimator shrinks the eigenvalues of the sample covariance toward zero.  \dele{In practice, the amount of shrinkage will depend on the particular subset chosen.}
	
\subsection{Calculation of Eigenvalues and Eigenvectors}

We now derive expressions for the eigenvalues and eigenvectors of the \Nystrom\ covariance estimator.  For notational convenience, given an $m \times n$ matrix $\matr{A}$ and a $k$-subset $I \subseteq \{1,\dotsc,m\}$, let $\matr{A}_{I\allin}$ denote the $k \times n$ submatrix formed by taking $k$ entire rows of $\matr{A}$ as indexed by $I$.  Also, given $r = \rank\pars{\matr{A}}$, we define the \term{thin singular value decomposition} (thin SVD) of $\matr{A}$ as $\matr{A} = \matr{U} \matr{D} \tp{\matr{V}}$, where $\matr{D}$ is an $r \times r$ diagonal matrix containing the nonzero singular values of $\matr{A}$, and $\matr{U}$ and $\matr{V}$ are $m \times r$ and $n \times r$ matrices whose columns are the corresponding left and right singular vectors.  

Using this notation, the eigenvalues and eigenvectors of the \Nystrom\ covariance estimator can be expressed as follows.


\begin{theorem}[Eigenvalues and eigenvectors of the \Nystrom\ covariance estimator] \label{thm:NystromEigenvalueComputation}
	Let $\matr{X}$ be a $p \times n$ matrix whose columns $\vect{x}_1,\dotsc,\vect{x}_n$ are i.i.d.\ random vectors such that \mbox{$\ev\parsi{\vect{x}_i} = \vect{0}$} and $\ev\parsi{\vect{x}_i\tp{\vect{x}}_i} = \matr{\Sigma}$ for $i = 1,\dotsc,n$.  
Given a $k$-subset \mbox{$I \subseteq \{1,\dotsc,p\}$}, define $J = \{1,\dotsc,n\} \setdiff I$, let \mbox{$r = \rank\parsii{\matr{X}_{I\allin}}$}, and let $\matr{X}_{I\allin}$ have the thin SVD $\matr{X}_{I\allin} = \matr{U}_X \matr{D}_X \tp{\matr{V}}_X$.  If $\matr{W}$ is the $p \times r$ matrix given by
\begin{align}
	\matr{W}_{I\allin} & = \frac{1}{\sqrt{n}}\,\matr{U}_X \matr{D}_X, \\
	\matr{W}_{J\allin} & = \frac{1}{\sqrt{n}}\,\matr{X}_{J\allin}\matr{V}_X,
\end{align}
then $\matr{W}\tp{\matr{W}} = \NystromCovEst$ is the \Nystrom\ covariance estimator of $\matr{\Sigma}$ given $I$, and its $r$ nonzero eigenvalues and corresponding eigenvectors are given by $\matr{\Lambda}^2$ and $\matr{U}$, where $\matr{W} = \matr{U}\matr{\Lambda}\tp{\matr{V}}$ is the thin SVD of $\matr{W}$.
\end{theorem}


\begin{IEEEproof}
	As in previous proofs, we let \mbox{$I = \{1,\dotsc,k\}$} and \mbox{$J = \{k+1,\dotsc,p\}$} without loss of generality.  Let $\matr{X}$ be partitioned as in \eqref{eq:XPartition}, and let $\matr{Y} = \matr{X}_{I\allin}$ have the thin spectral decomposition $\matr{Y} = \matr{U}_Y \matr{D}_Y \tp{\matr{V}}_Y$.  Letting
\begin{equation}
	\matr{W} = \frac{1}{\sqrt{n}}\matrels{\matr{U}_Y \matr{D}_Y \\ \matr{Z} \matr{V}_Y},
\end{equation}
we have
\begin{align}
	\matr{W}\tp{\matr{W}}
	& = \frac{1}{n}\matrels{\matr{U}_Y \matr{D}_Y \\ \matr{Z} \matr{V}_Y}
		\matrels{\matr{D}_Y\tp{\matr{U}}_Y  & \tp{\matr{V}}_Y \tp{\matr{Z}}} \\
	& = \frac{1}{n}\matrels{\matr{U}_Y \matr{D}^2_Y \tp{\matr{U}}_Y & 
		\matr{U}_Y \matr{D}_Y \tp{\matr{V}}_Y \tp{\matr{Z}} \\
		\matr{Z} \matr{V}_Y \matr{D}_Y \tp{\matr{U}}_Y & 
		\matr{Z} \matr{V}_Y \tp{\matr{V}}_Y \tp{\matr{Z}}} \\
	& = \frac{1}{n}\matrels{\matr{Y}\tp{\matr{Y}} & \matr{Y}\tp{\matr{Z}} \\
		\matr{Z} \tp{\matr{Y}} & \matr{Z} \matr{V}_Y \tp{\matr{V}}_Y \tp{\matr{Z}}}.
\end{align}
Noting that the Moore-Penrose pseudoinverse of $\tp{\matr{Y}}$ is
\begin{equation}
	\pinv{\parsii{\tp{\matr{Y}}}} = \matr{V}_Y \inv{\matr{D}}_Y \tp{\matr{U}},
\end{equation}
we see that \Nystrom\ projection of \eqref{eq:NystromProjection} is equivalent to
\begin{equation}
	\matr{P} \equiv \tp{\matr{Y}}\pinv{\parsii{\tp{\matr{Y}}}}
		= \matr{V}_Y \matr{D}_Y \tp{\matr{U}}_Y \matr{U}_Y \inv{\matr{D}}_Y \tp{\matr{V}}_Y
		= \matr{V}_Y \tp{\matr{V}}_Y,
\end{equation}
and thus $\matr{W}\tp{\matr{W}} = \NystromCovEst$.  Consequently, if $\matr{U}\matr{\Lambda}\tp{\matr{V}}$ is the thin SVD of $\matr{W}$, then
\begin{equation}
	\NystromCovEst = \matr{U}\matr{\Lambda}\tp{\matr{V}}\matr{V}\matr{\Lambda}\tp{\matr{U}}
		= \matr{U}\matr{\Lambda}^2\tp{\matr{U}}
\end{equation}
is the (thin) spectral decomposition of $\NystromCovEst$, where $\matr{\Lambda}^2$ is an $r \times r$ matrix containing the nonzero eigenvalues of $\NystromCovEst$, and $\matr{U}$ is a $p \times r$ matrix whose columns are the corresponding eigenvectors.
\end{IEEEproof}


For fixed $k$, the computational complexity of the spectral decomposition of $\NystromCovEst$ is dominated by two operations.  The first is the multiplication $\matr{X}_{J\allin} \matr{V}_X$, the cost of which scales as $\bigO(p\,n)$; the second is the thin SVD of $\matr{W}$, the cost of which scales as $\bigO(p)$.  Thus, for $n$ fixed the overall cost of eigenanalysis scales linearly in the dimension $p$, making \Nystrom\ covariance estimator an appealing choice in extremely high-dimensional settings.

\section{Example Application: Adaptive Beamforming} \label{sec:Beamforming}

We conclude our discussion with two examples of the use of \Nystrom\ covariance estimation in practical applications.  For the first example, we examine the classical signal processing problem of \term{beamforming} \cite{Brennan1973,VanVeen1988,Feldman1994,Bell2000}, which involves tailoring the signal response (or ``beam pattern'') of an array of receiving elements to enhance one's ability to detect a desired signal.

\subsection{Narrowband Beamforming Model}

\begin{figure}
	\centering
		\includegraphics[width=0.4\textwidth]{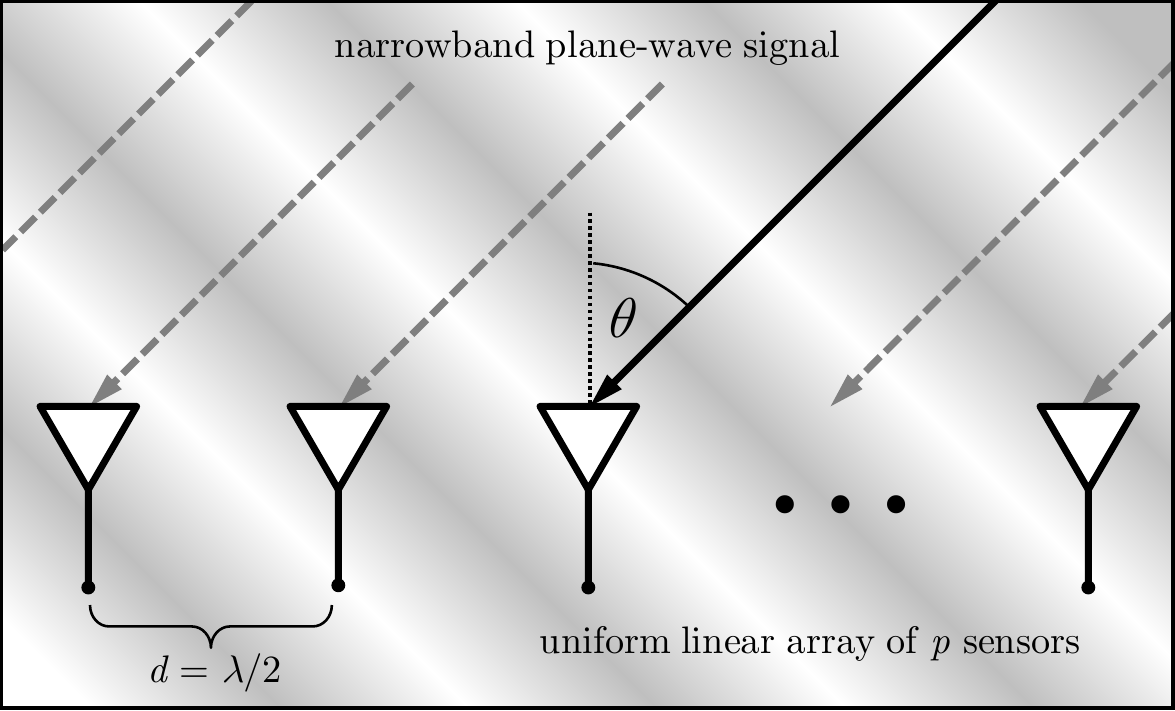}
	\caption{A simple narrowband beamforming model.}
	\label{fig:ArrayModel}
\end{figure}

We adopt a standard beamforming model, illustrated in \figref{fig:ArrayModel}.  Consider a collection of $p$ sensing elements, which sample an incoming plane-wave signal at discrete time intervals.  We assume the signal of interest is narrowband \dele{(i.e.\ its frequency content is restricted to a small interval around a given carrier frequency $f_c$)}and that the sensors are placed in a straight line with equal spacing (known as a \term{uniform linear array}).  To avoid aliasing in the spatial sampling of the signal, we assume an element spacing of $d = \lambda_c/2$, where $\lambda_c$ is the carrier wavelength.  In addition, let $\theta$ denote the angle between the wave's direction of propagation and a vector normal to the array, referred to as the \term{angle of arrival}.  


\dele{We represent the signal of interest by its \term{complex envelope}, a complex-valued function that characterizes the signal's deviation from the amplitude and phase of the carrier tone.  }

Assume that there are $k \leq p$ incoming signals, and let $z_i(t) \in \complex$ denote the complex envelope of the $i$-th signal at the $t$-th sample time, for $i \in \{1,\dotsc,k\}$ and \mbox{$t \in \{1,2,\dotsc\}$}.  Arranging the signal response for all array elements into a 
vector, the total received signal at the $t$-th sample time (referred to as a ``snapshot'') is 
\begin{equation}  \label{eq:BeamformingModel}
	\vect{x}(t) = \sum_{i=1}^{k} \vect{a}(\theta_i)\,z_i(t) + \vect{n}(t),
\end{equation}
where $\vect{n}(t) \in \complex^p$ is additive noise and where $\vect{a}(\theta_i) \in \complex^p$ represents the amplitude change and phase delay at each sensor as a function of $\theta_i$, the angle of arrival of the $i$-th signal.  Assuming a constant amplitude response across all elements and letting the phase response at the first element be zero, \dele{the time delay between successive elements is $\sin(\theta_i)/(2f_c)$, and} we have
\begin{equation}  \label{eq:ArrayResponse}
	\bracks{\vect{a}(\theta_i)}_l = e^{-j\pi (l-1) \sin(\theta_i)},
\end{equation}
for $l = 1,\dotsc,p$ and where $j = \sqrt{-1}$.

Now, let us assume that out of the $k$ incoming signals, only one---say, $z_1(t)$---is of interest to us, and all others are considered interference.  In this context, the classical narrowband beamforming problem can be stated as follows: given a collection of $n$ snapshots $\{\vect{x}(1),\dotsc,\vect{x}(n)\}$, determine a weight vector $\vect{w} \in \complex^p$ such that the output of the linear filter (or ``beamformer'')
\begin{equation}
	\hat{z}_1(t) \equiv \herm{\vect{w}}\vect{x}(t)
\end{equation}
such that the mean squared error of $z_1(t) - \hat{z}_1(t)$ is small. 

For our simple example, let us assume that given the narrowband beamforming model in \eqref{eq:BeamformingModel}, the signals $z_1(t), \dotsc, z_m(t)$ and the noise $\vect{n}(t)$ are stationary zero-mean Gaussian random processes that are statistically independent across time samples.  Let $\sigma_i^2 = \ev\parsii{z_i^2(t)}$ denote the expected power of the $i$-th signal, and let $\sigma^2_n = \ev\parsii{n^2(t)}$ denote the expected noise power.  In this case, it can be shown \cite{VanTrees1968,Bell2000} that the optimal beamformer (in the sense of minimum MSE) is given by
\begin{equation} \label{eq:OptimalBeamformer}
	\vect{w}_{\text{opt}} \equiv \min_{\vect{w} \in \complex^p}
		\ev\pars{\herm{z_1(t) - \vect{w}}\vect{x}(t)}^2
		= \matr{\Sigma}^{-1} \vect{a}\!\pars{\theta_1}\,\sigma^2_1,
\end{equation}
where
\begin{equation}
	\matr{\Sigma} = \ev\pars{\vect{x}(t)\,\herm{\vect{x}}(t)}
		= \sum_{i=1}^k \sigma_i^2\,\vect{a}\!\pars{\theta_i}\herm{\vect{a}}\!\pars{\theta_i}
		+ \sigma_n^2\,\idmat_p\,.
\end{equation}
Unfortunately, there are a number of barriers to realizing this optimal beamformer in practice.  Even if our modeling assumptions hold, we need to know the power $\sigma^2_1$ and angle of arrival $\theta_1$ of the signal of interest, as well as the covariances of the interference and noise.

\subsection{Beamforming Using the \Nystrom\ Covariance Estimator}

Let assume that $\sigma^2_1$ and $\theta_1$ are known.  In this case, we might consider approximating $\matr{\Sigma}$ by the sample covariance
\begin{equation}
	\matr{S} = \frac{1}{n} \sum_{t=1}^n \vect{x}(t)\,\herm{\vect{x}}(t),
\end{equation}
given a collection of $n$ observations.  Although this approach will recover the optimal beamformer as $n \to \infty$, in practice the number of samples is bounded, and any assumption of stationarity regarding the signal and interference usually is valid only over a limited time window.  This issue is especially problematic when the number of elements $p$ is large, as we may not have enough samples for $\matr{S}$ to be well-conditioned (or even invertible).  As a result, direct substitution of the sample covariance into \eqref{eq:OptimalBeamformer} is rarely an acceptable solution.

One alternative is to replace $\matr{S}$ with a low-rank approximation \cite{Feldman1994,Parker2005,Spendley2008,Zulch1998}.  For example, consider the optimal rank-$m$ approximation
\begin{equation}
	\matr{S}^{\ast}_k = \matr{U}_k\matr{\Lambda}_k\herm{\matr{U}}_k,
\end{equation}
where $\matr{\Lambda}_k$ is a $k \times k$ diagonal matrix containing the $k$ largest eigenvalues of $\widehat{\matr{\Sigma}}$, and $\matr{U}_k$ is a $p \times k$ matrix containing the corresponding eigenvectors.  We can define a ``projection beamformer'' (denoted $\vect{w}_{\text{proj}}$) by approximating \eqref{eq:OptimalBeamformer} as
\begin{equation} \label{eq:ProjectionBeamformer}
	\vect{w}_{\text{proj}}
		\equiv \pinv{\parsii{\matr{S}^{\ast}_k}}
		\vect{a}(\theta_1)\,\sigma^2_1,
\end{equation}
where 
\begin{equation}
	\pinv{\parsii{\matr{S}^{\ast}_k}}
		= \matr{U}_k \inv{\matr{\Lambda}}_k \herm{\matr{U}}_k.
\end{equation}
Consequently, the filter output can be expressed as
\begin{equation}
	\hat{z}_1(t) = \herm{\vect{w}}_{\text{proj}}\,\vect{x}(t)
		= \sigma^2_1 \herm{\vect{a}}(\theta_1)\,
		\pinv{\parsii{\matr{S}^{\ast}_k}} \matr{P} \vect{x}(t),
\end{equation}
where $\matr{P} \equiv \matr{U}_m \herm{\matr{U}}_m$ is an orthogonal projection onto the $m$-dimensional principal subspace of $\matr{S}$.  Thus, the projection beamformer maps the incoming signal onto a low-rank subspace, and then approximates the behavior of optimal beamformer within this space.  If the range of $\matr{P}$ is close to the span of $\{\vect{a}(\theta_1) z_1(t), \dotsc, \vect{a}(\theta_m) z_m(t)\}$, any signal power lost in the low-rank projection will be largely due to noise, thus improving overall estimation performance.  Of course, the effectiveness of this approach in practice depends on the powers of the signal and interference relative to the noise, as well as the sample size.

The projection beamformer of \eqref{eq:ProjectionBeamformer} presents an excellent opportunity to apply the \Nystrom\ covariance estimator.  Substituting $\NystromCovEst$ for the optimal low-rank approximation of the sample covariance, we define the ``\Nystrom\ beamformer'' (denoted $\vect{w}_{\text{Nyst}}$) as
\begin{equation} \label{eq:NystromBeamformer}
	\vect{w}_{\text{Nyst}}
		\equiv \pinv{\parsii{\NystromCovEst}}\,\vect{a}(\theta_1)\,\sigma^2_1,
\end{equation}
given a $k$-subset $I \subseteq \{1,\dotsc,p\}$.  In contrast to to the projection beamformer, the \Nystrom\ beamformer characterizes the signal covariance only over a subset of $k \leq p$ sensors in the array; the rest of the covariance is inferred as a function of observed correlations between elements in this subset and the remaining sensors.  The goal of this approach is to achieve performance comparable to that of the projection beamformer, while realizing significant reductions in computational cost.

\subsection{Experimental Results}

To compare the performance of various beamforming approaches, we simulated a uniform linear array with $p = 100$.  We considered a case with $k = 7$ signals, where the angle of arrival of desired signal was $10$ degrees, and the six interference signals had angles of arrival of $-65$, $-30$, $-25$, $30$, $45$, and $60$ degrees.  For all experiments, we assumed a constant noise power of $\sigma^2_n = 1$ and a constant interference-to-noise ratio (INR) of 20~dB.

Given a signal-to-noise ratio (SNR) for the desired signal, we studied the performance of each method as a function of the number of snapshots $n$.  Our primary measure of performance was the \term{signal to interference and noise ratio} (SINR) of the estimated signal, defined as
\begin{equation} 
	\text{SINR}
		= \frac{\sum_{t=1}^n \abs{\hat{z}_1(t)}^2}
		{\sum_{t=1}^n \abs{\herm{\vect{w}}\vect{z}(t)}^2}\,,
\end{equation}
where $\hat{z}_1(t) = \herm{\vect{w}}\vect{x}(t)$ and $\vect{z}(t)$ is the sum of the received interference and noise at the $t$-th sample time,
\begin{equation}  
	\vect{z}(t) \equiv \sum_{i=2}^{m} \vect{a}(\theta_i) z_i(t) + \vect{n}(t).
\end{equation}


\begin{figure}[t]
	\centering
	\includegraphics[width=0.5\textwidth]{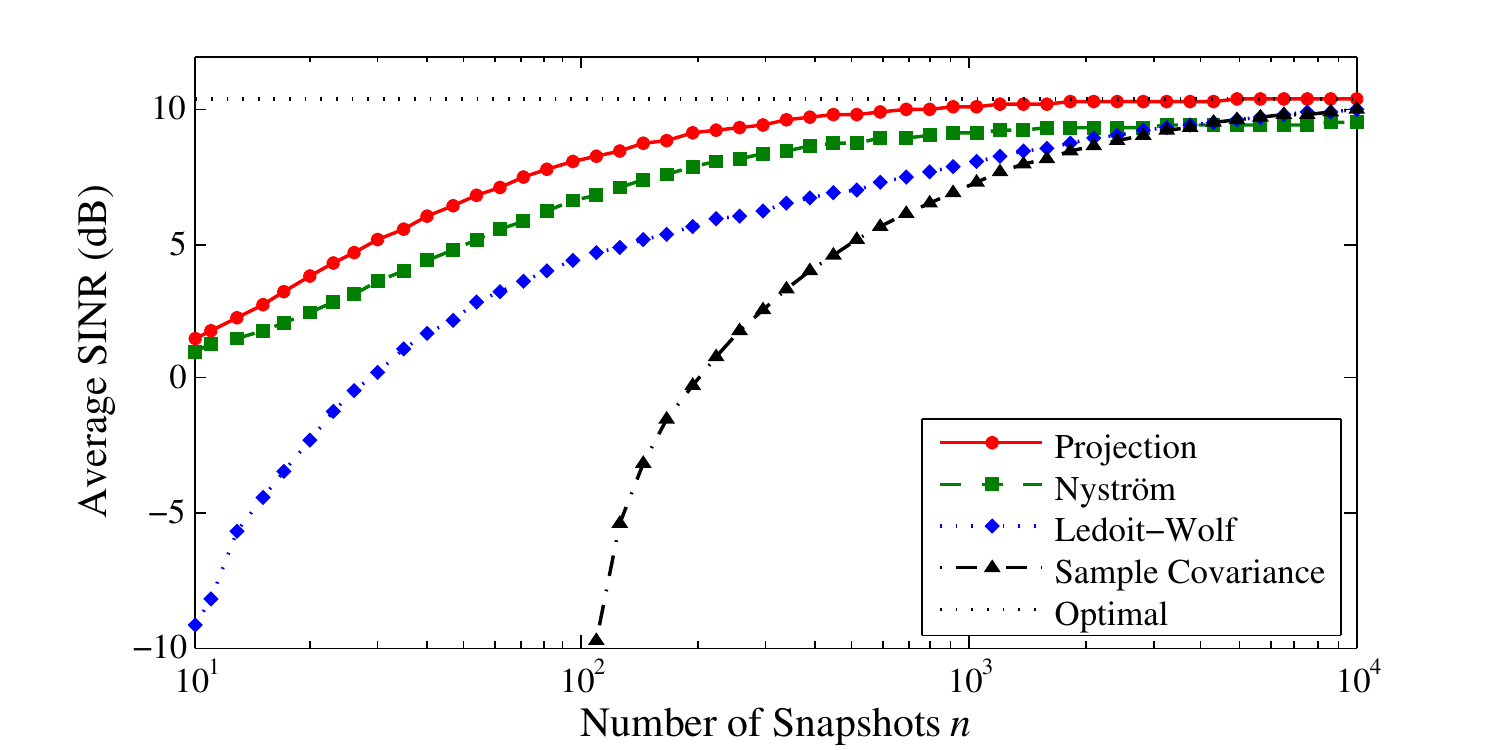} \\
	\includegraphics[width=0.5\textwidth]{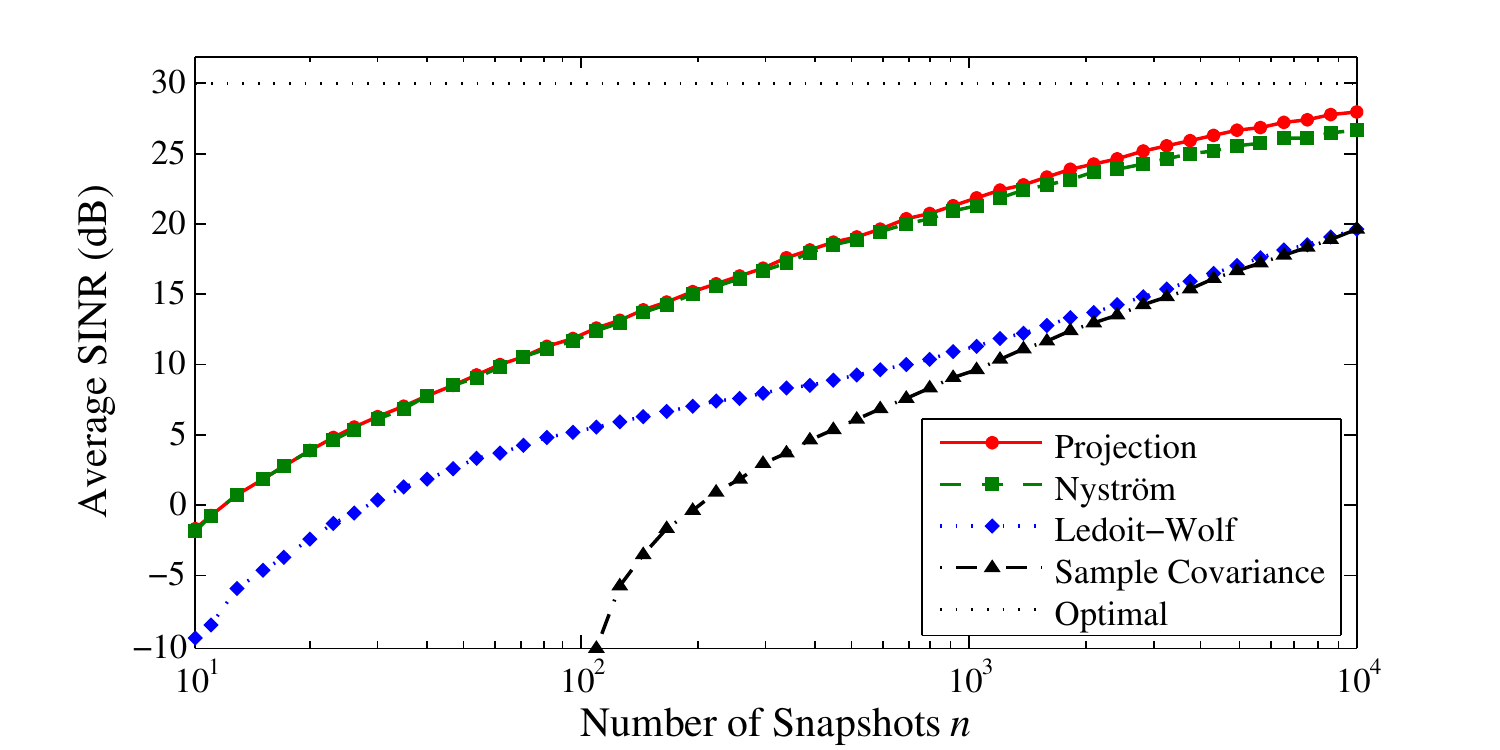} \\
	\includegraphics[width=0.5\textwidth]{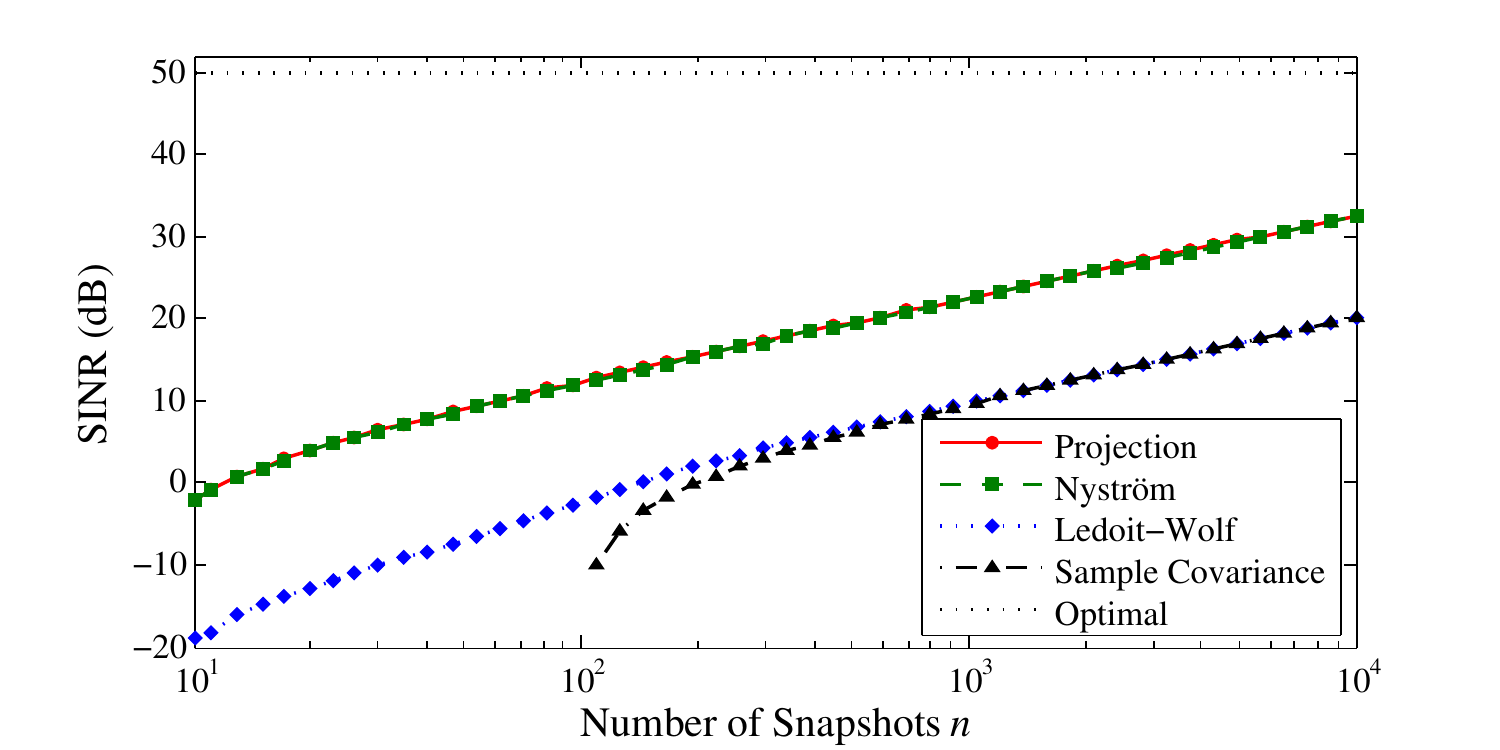}
	\caption{SINR as a function of number of snapshots for various beamforming approaches, given an INR of 20~dB and SNR values of $-10$~dB (top), 10~dB (middle), and 30~dB (bottom).}
	\label{fig:BeamformingExamples}
\end{figure}


\figref{fig:BeamformingExamples} shows the performance of various beamformers for three different SNR levels ($-10$~dB, $10$~dB, and $30$~dB).  For each beamformer, we plot the SINR in dB as a function of the number of snapshots, averaged over 1000 experimental trials.  Note that the horizontal axis is measured on a logarithmic scale, ranging from $n = 10$ to $n = 10^4$ samples.   

Results are shown for projection beamformer of \eqref{eq:ProjectionBeamformer} and the \Nystrom\ beamformer of \eqref{eq:NystromBeamformer}, where the latter is computed using a uniformly random $k$-subset $I \subseteq \{1,\dotsc,p\}$.  For comparison, we include two approximations of the optimal beamformer: one where we have replaced $\matr{\Sigma}$ in \eqref{eq:OptimalBeamformer} with the sample covariance, and another where we have substituted the Ledoit-Wolf covariance estimator of \cite{Ledoit2004}.  We also show an upper bound given by the theoretical SINR of the optimal beamformer,
\begin{equation} 
	\text{SINR}_{\text{opt}}
		= \frac{\ev\abs{\herm{\vect{w}}_{\text{opt}}\vect{x}(t)}^2}
		{\ev\abs{\herm{\vect{w}}_{\text{opt}}\vect{z}(t)}^2}
		= \frac{\herm{\vect{w}}_{\text{opt}} \matr{\Sigma} \vect{w}_{\text{opt}}}
		{\herm{\vect{w}}_{\text{opt}} \matr{\Sigma}_z \vect{w}_{\text{opt}}}\,,
\end{equation}
where $\matr{\Sigma}_z = \ev\parsii{\vect{z}(t)\,\herm{\vect{z}}(t)}$ is the covariance of the interference plus noise.

\begin{figure}
	\centering
	\includegraphics[width=0.5\textwidth]{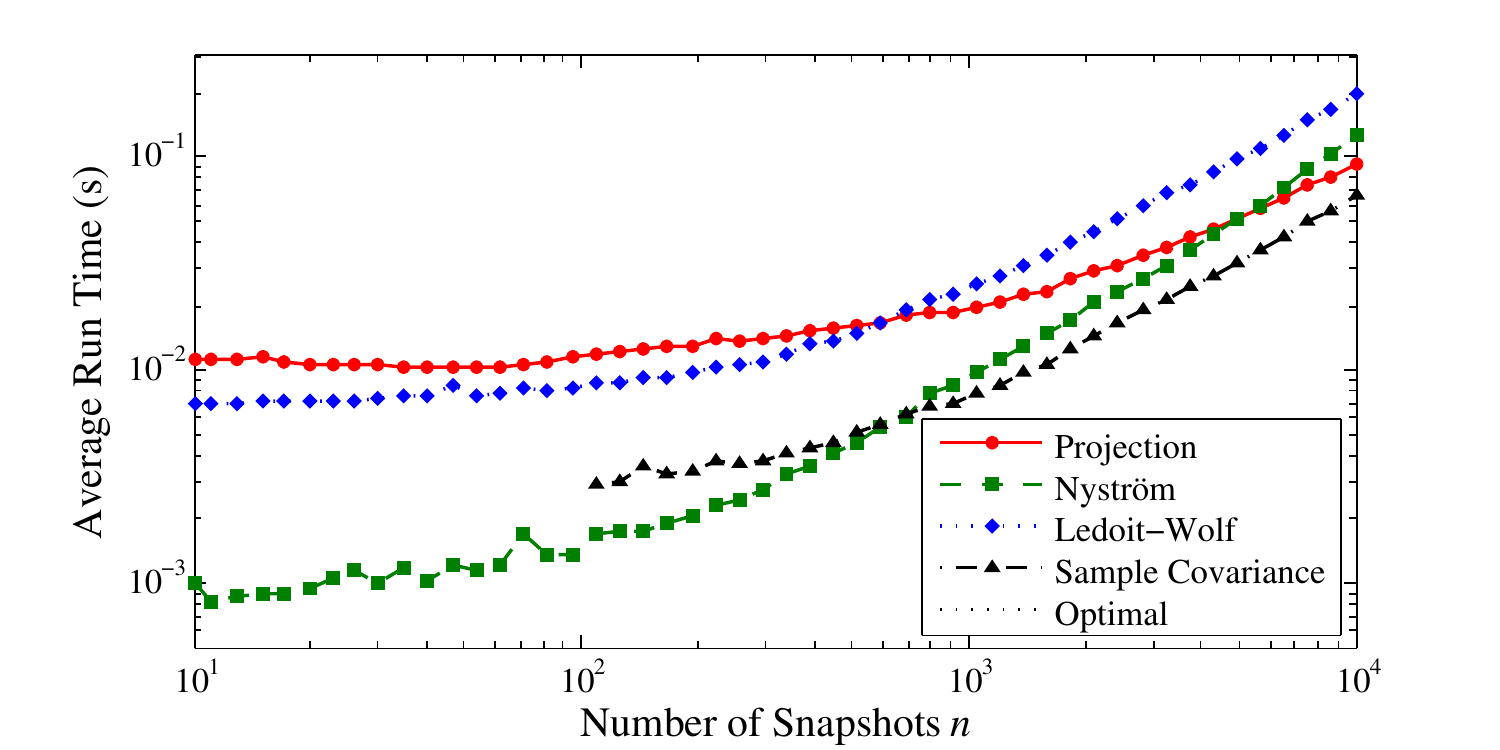}
	\caption{Average running time in seconds of various beamformers for INR = 20~dB and SNR = $-10$~dB, as a function of the number of snapshots.}
	\label{fig:BeamformingRunTimes}
\end{figure}

Since the purpose of the \Nystrom\ beamformer is to achieve satisfactory error performance for a low computational cost, we also compared the running time of each algorithm.  For the low-SNR case ($-10$~dB), \figref{fig:BeamformingRunTimes} shows the average CPU time required to compute each beamformer on a personal computer with an Intel dual-core 2.33-GHz processor, as a function of the number of snapshots.  


In the low-SNR case, the SINR performance of the \Nystrom\ beamformer is comparable to that of the projection beamformer, with the former trailing the latter by a margin of 0.4 to 1.6~dB.  Both low-rank methods perform considerably better than beamforming using the sample and Ledoit-Wolf covariance estimators.  We see that the sample covariance beamformer (which is undefined until the sample covariance becomes invertible at $n=100$ snapshots) exhibits the poorest performance, due to the over-dispersion of its eigenvalues.  Although shrinkage provided by the Ledoit-Wolf beamformer does improve the conditioning of the sample covariance beamformer, both approaches remain inferior to the low-rank methods until the number of snapshots grows large ($n > 4000$).

We observe similar performance trends in the medium-SNR ($10$~dB) and high-SNR ($30$~dB) cases.  At an SNR of 10 dB, the SINR performance of the \Nystrom\ beamformer lags behind that of the projection beamformer by a margin of 0.06 to 1.4~dB, and at an SNR of 30 dB, the difference is minor (less than $0.15$ dB) across all values of $n$.  In both cases, the SINR performance of the Ledoit-Wolf and sample covariance estimators is consistently about 10 to 20 ~dB less than that of the low-rank methods.

In terms of computation, the \Nystrom\ beamformer requires about an order of magnitude less time to run than the projection and Ledoit-Wolf methods, for $n$ up to around 100.  However, as $n$ grows large with $p$ fixed, the complexity of all four approaches is dominated by the $\bigO(n)$ cost of computing covariance terms over the set of snapshots, and thus the computational differences between the beamformers become less dramatic.  Alternatively, if we were to let $p$ grow while keeping $n$ fixed, the computational cost would instead be dominated by the low-rank approximation step (or by matrix inversion, in the case of the Ledoit-Wolf or sample covariance beamformers).  In this case, the \Nystrom\ beamformer would continue to exhibit significant computational savings when compared to the other methods.

\section{Example Application: Image Denoising} \label{sec:Denoising}

For our second example of an application for \Nystrom\ covariance estimation, we consider the problem of \term{image denoising} \cite{Muresan2003,Elad2006,Dabov2007}.  Let $\vect{z} \in \reals^m$ represent an 8-bit grayscale image with $m$ pixels, where each element $z_i \in \{0,\dotsc,255\}$ is the intensity of the $i$-th image pixel for $i = 1,\dotsc,m$.  Assume that we obtain a noisy version of this image $\vect{x} = \vect{z} + \vect{n}$, where $\vect{n} \distas \normpdf_m\pars{\vect{0},\sigma^2\idmat_m}$.  Given $\vect{x}$, we wish to compute an estimate $\hat{\vect{z}}$ of the clean image $\vect{z}$.  

Many approaches to image denoising involve computing local decompositions of the noisy signal over sets of nearby pixels (or ``image patches'').  We will investigate a denoising solution based on principal components analysis (PCA) of groups of patches, which requires estimating and then decomposing local covariance matrices.  \dele{Although the processing of individual covariance matrices will not present a computational challenge, a typical image may comprise a large number of patches.  Consequently, these computations will need to be repeated many times, and we will be able to achieve computational gains by performing PCA using the \Nystrom\ covariance estimator in lieu of the sample covariance.}


\begin{figure*}[t]
	\centering
	\includegraphics[width=2.9in, bb=0 0 512 342]{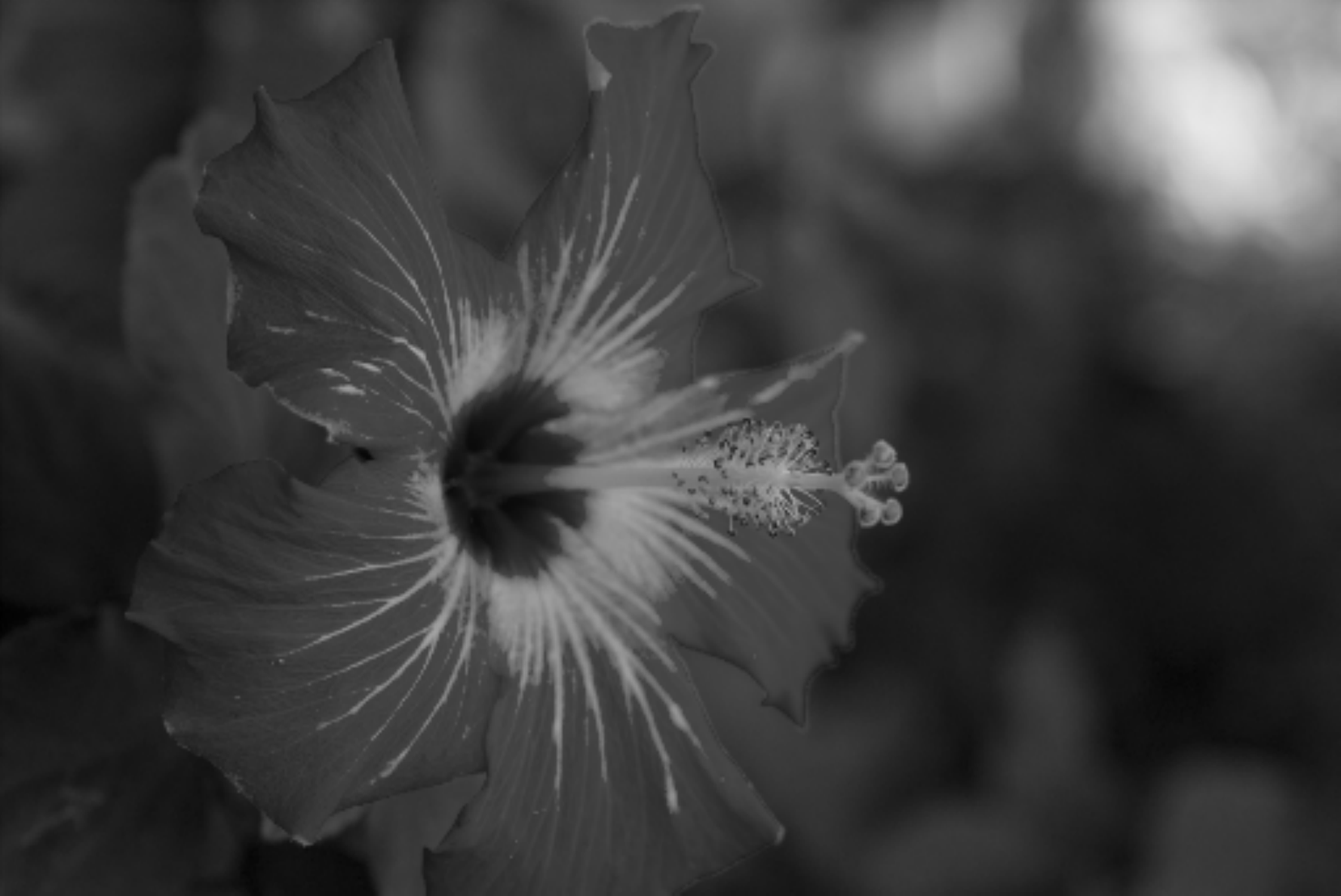}
	\hspace{0.05in}
	\includegraphics[width=2.9in, bb=0 0 512 342]{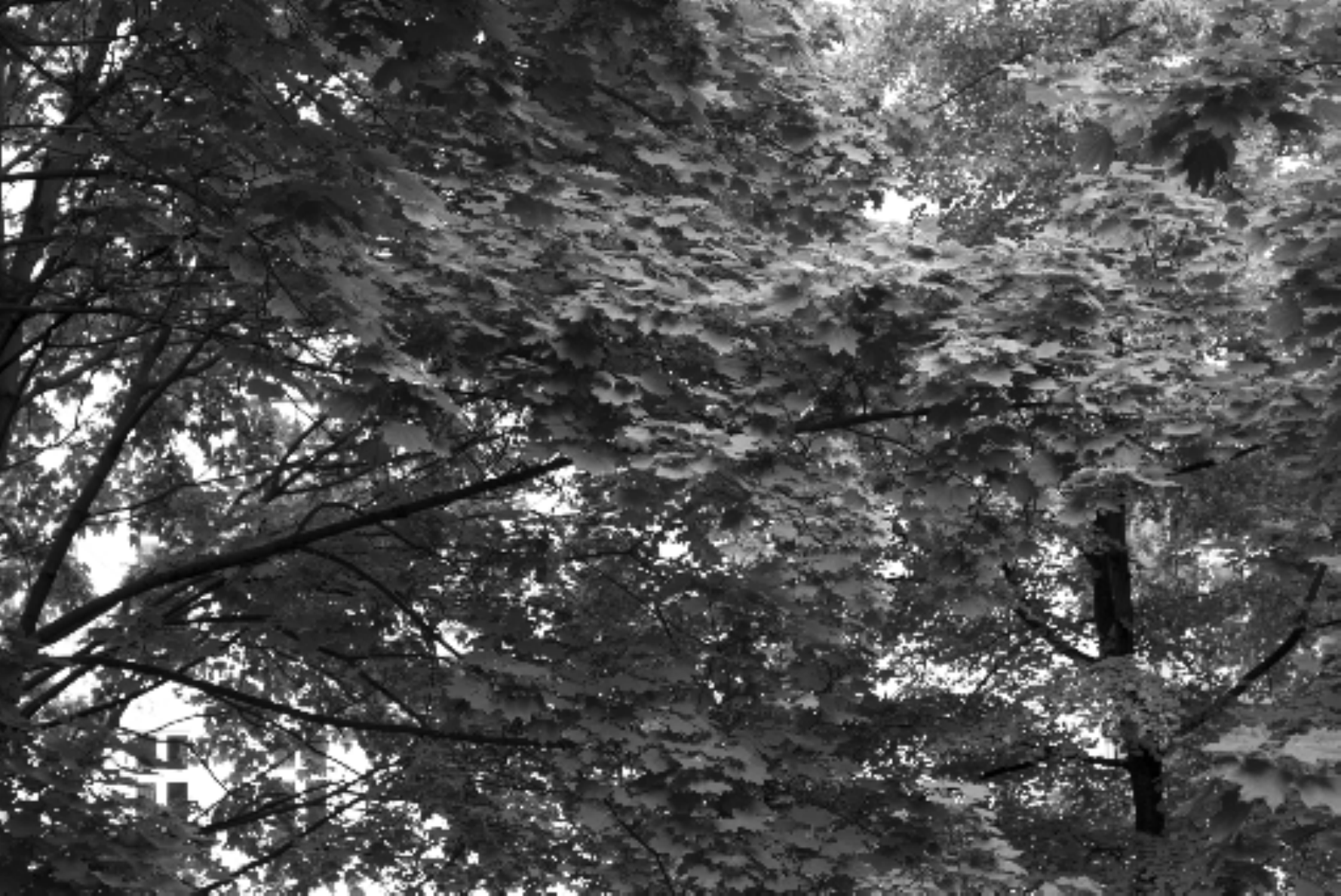}
	\\ \vspace{0.13in}
	\includegraphics[width=2.9in, bb=0 0 512 342]{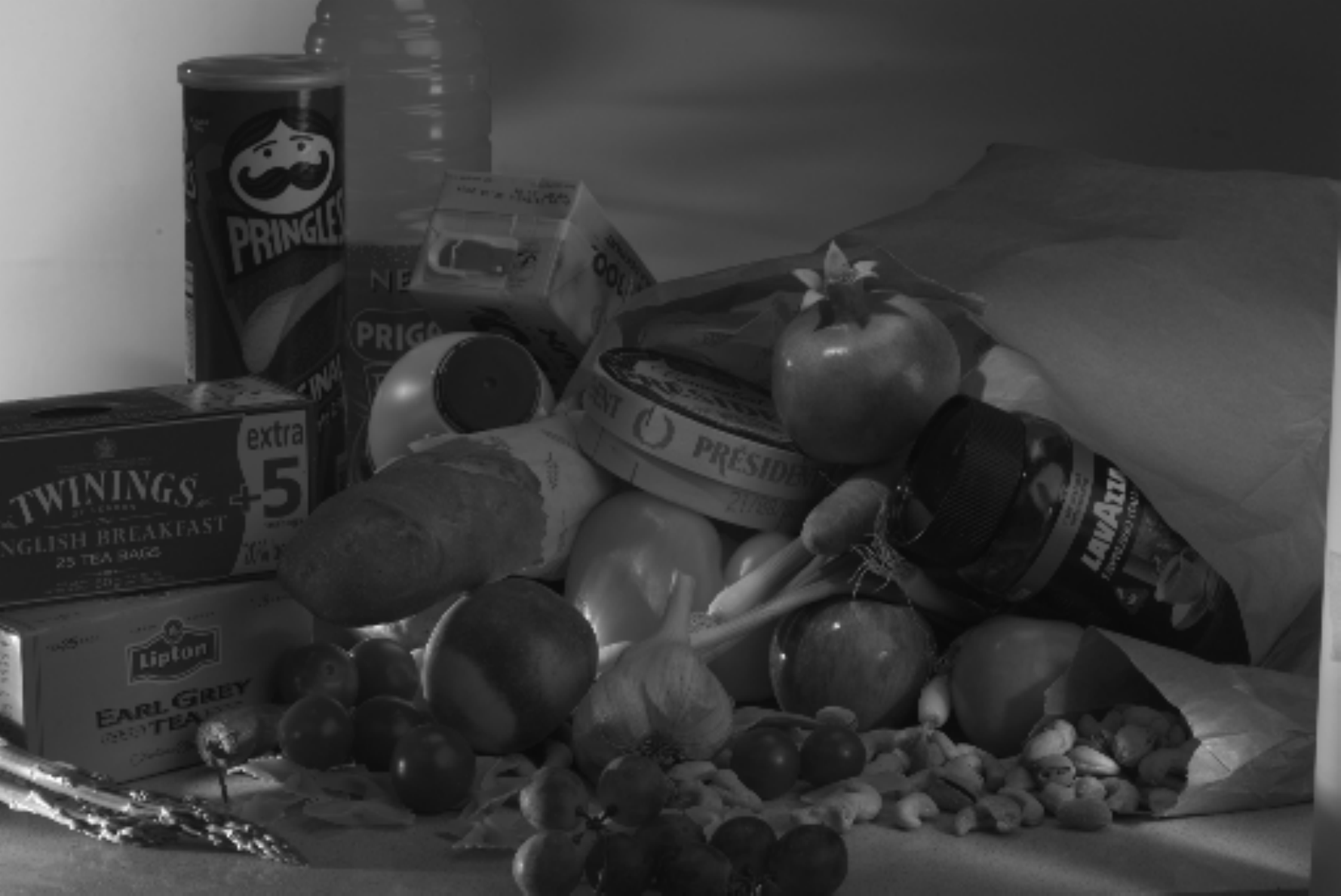}
	\hspace{0.05in}
	\includegraphics[width=2.9in, bb=0 0 512 342]{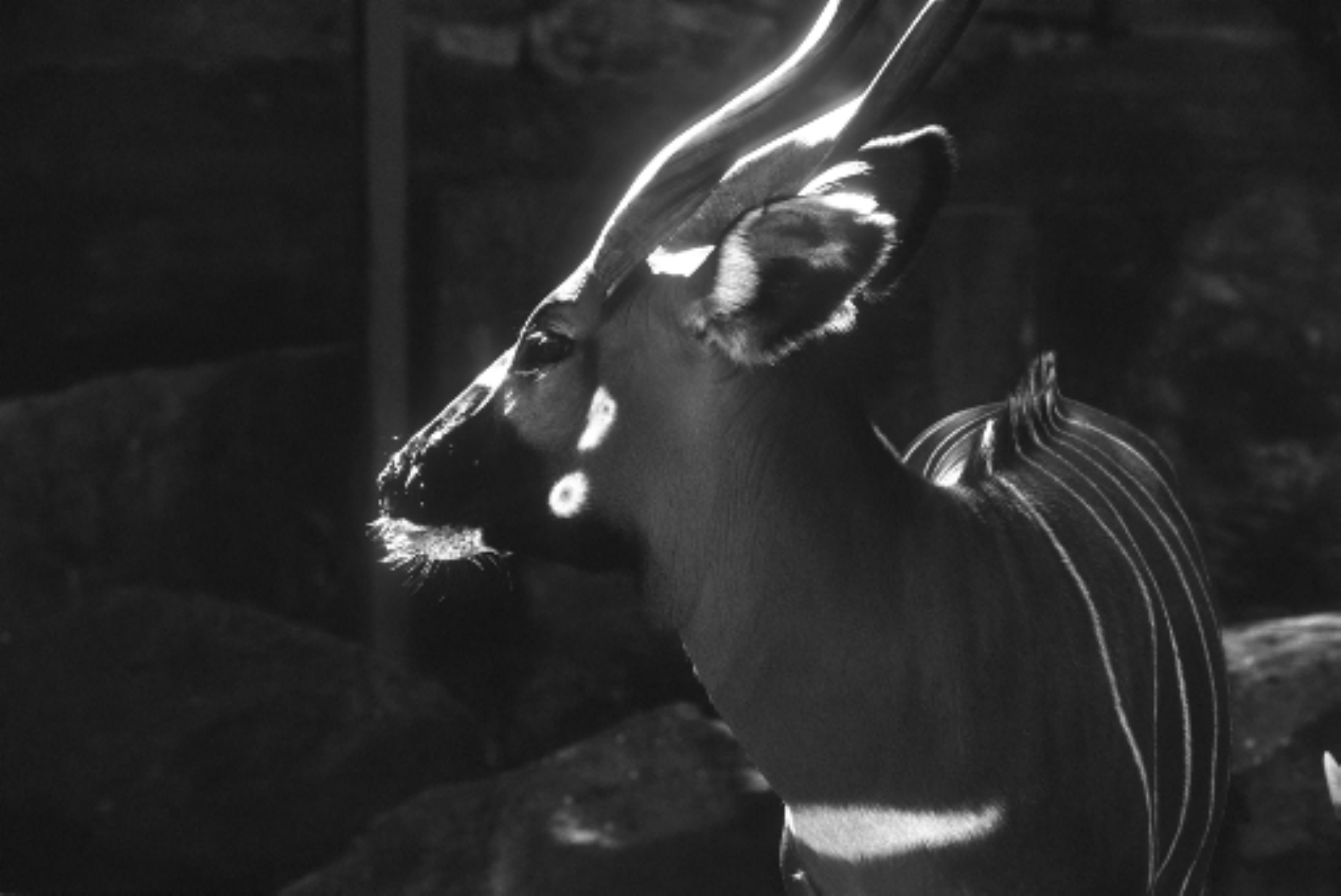}
	\caption{High-resolution test images for image denoising experiments.  Top left: \textit{flower} ($2256 \times 1504$). Top right: \textit{leaves} ($3008 \times 2000$).  Bottom left: \textit{hdr} ($3072 \times 2048$).  Bottom right: \textit{deer} ($4032 \times 2640$).} 
	\label{fig:TestImages}
\end{figure*}


\subsection{Subspace Estimation for Image Denoising} \label{sec:SubspaceModel}

We begin by developing a simple image model allowing us to perform denoising as a subspace estimation problem.  Since natural images often possess a high degree of local similarity, a common assumption in image processing is that given a spatially proximate set of pixels, there exists some transformation under which these pixels admit a sparse representation \cite{Mika1999,Muresan2003}.  Let us partition $\vect{z}$ into $q$ sub-vectors $\vect{z}_1,\dotsc,\vect{z}_q$, where each $\vect{z}_i \in \reals^p$ is a local set of $p$ pixels (referred to as a ``patch''), with $m = p\,q$.  The observed image patches are $\vect{x}_i = \vect{z}_i + \vect{n}_i$, where $\vect{n}_i \distas \normpdf_p\pars{\vect{0},\sigma^2\idmat_p}$ for $i = 1,\dotsc,q$.  Assume now that each patch $\vect{z}_i$ is restricted to a subspace of dimension $k_i \leq p$, which we denote $S_i$.  In this case, it can be shown \cite{Scharf1991} that a linear least-squares estimator of $\vect{z}_i$ given $\vect{x}_i$ is 
\begin{equation} \label{eq:LLSEDenoiser}
	\hat{\vect{z}}_i \equiv \matr{P}_i \matr{x}_i
		= \matr{z}_i + \matr{P}_i \matr{n}_i\,
\end{equation}
where $\matr{P}_i$ represents an orthogonal projection onto $S_i$.  This estimator preserves $\matr{z}_i$, while removing all noise except for the component in the signal subspace.

In practical applications, the subspaces $S_1,\dotsc,S_q$ typically are not known and must be estimated from the noisy image.  Consider a set of $n$ noisy patches $\{\vect{x}_i \st i \in I\}$, which lie within a region of the image defined by a set of $n$ patch indices $I \subseteq \{1,\dotsc,q\}$.  If the patches in this region have similar signal characteristics\change{---perhaps due to spatial proximity, or because they represent similar objects or textures---}{, }then we may assume that all $S_i$ are equal for $i \in I$.  This assumption allows us to estimate the subspaces from the $k_i$ principal components of the sample covariance matrix
 \begin{equation} \label{eq:ImageSampleCovariance}
	\matr{S} = \frac{1}{n}\sum_{i \in I} \vect{x}_i \tp{\vect{x}}_i.
\end{equation}
Note that in practice the subspace dimension is also unknown and may vary across regions.  To address this problem, one may attempt to solve the rank estimation problem of determining $k_i$ from the noisy image.  However, for our simple example we set $k_i$ to a fixed value across all images.  

By repeating the component analysis for all regions, we can obtain a full set of projection matrices for computing the image estimate.  We will refer to image estimation as in \eqref{eq:LLSEDenoiser} where the projections are estimated from the principal components of sample covariances as the \term{PCA image denoiser}.

Depending on the size of the image, estimating the full set of orthogonal projections for PCA denoising can require computing and then decomposing or inverting a large number of covariance matrices.  Consequently, we may realize significant improvements in computation by replacing the sample covariance with a \Nystrom\ covariance estimate, which we will refer to as the \term{\Nystrom\ image denoiser}.  \dele{To perform rank estimation for the \Nystrom\ image denoiser, we first specify a maximum subspace rank (equal to the order of the \Nystrom\ approximation), and then retain only those components needed to achieve the necessary fraction of the total spectral power.}

Although we have defined both approaches for disjoint image patches and regions, one may want to allow for some amount of overlap among these sets.  This modification increases the number of available samples, while also mitigating some of the artifacts that occur at patch boundaries.  While commonly used in practice \cite{Muresan2003, Elad2006}, note that allowing for overlapping patches conflicts with independence assumptions of our additive noise model.  It also increases computation due to the additional number of patches, and because estimated patches may need to be reweighted before they are combined.

\subsection{Experimental Results}

To compare the PCA and \Nystrom\ image denoisers, we examined their performance when applied to a selection of 8-bit high-resolution test images from \cite{Garg2010}.  The four images used are shown in \figref{fig:TestImages}.  Our primary measure of performance is the \term{peak signal-to-noise ratio} (PSNR) of the estimated image, a standard metric used throughout the image processing literature.  The PSNR of the denoised image $\hat{\vect{z}}$ is defined as
\begin{equation} \label{eq:PSNR}
	\text{PSNR}
		= \frac{z_{\max}^2}{\norm{\hat{\vect{z}} - \vect{z}}^2}\,,
\end{equation}
where $z_{\max}$ denotes the maximum allowable pixel value (in our case, 255).  After generating noisy versions of each image for noise levels of $\sigma = 10$, $20$, and $50$, we attempted to reconstruct the original image using the PCA and \Nystrom\ approaches, assuming a fixed subspace dimension of $k = 4$.

By examining denoising results for different patch and region sizes over the selection of test images, we determined a set of default parameter values that yielded a reasonable balance of performance and computation.  For both algorithms, we divided the image into regions of $32 \times 32$ pixels, with adjacent regions having 50\% overlap.  We then divided each region into $8 \times 8$ patches, also with 50\% overlap.  Thus, each region contained $n = 49$ patches, which were used to estimate a local covariance of dimension $p = 64$.  

In the case of the \Nystrom\ denoiser, covariance estimation was performed for each region using the \Nystrom\ covariance estimator, conditioned on a set of $k$ patch vectors chosen uniformly at random. Once the $k$ principal components and the estimated projection were computed, the denoised patches were superimposed to reconstruct an estimate of the original image.

As a benchmark, we provide results from two other patch-based denoising methods: the K-SVD algorithm of \cite{Elad2006} and the BM3D algorithm of \cite{Dabov2007}.  The first algorithm performs denoising based on a trained ``dictionary'' of components obtained from the noisy image, while the second jointly filters groups of image patches by arranging them into 3-D arrays.  For both algorithms, denoising was performed using code from the authors' respective websites, with most parameters set to their default values.  The only parameters we adjusted were the dictionary size and maximum number of training blocks for the K-SVD algorithm; to accommodate the high resolution of test images, these values were increased to 1024 and 130,000, respectively.

Results for all four algorithms are listed in \tabref{tab:DenoisingHiRes}.  For each test image, algorithm, and noise level, we list the average empirical PSNR over 10 realizations of each noisy image.  We see that the performance of the PCA and \Nystrom\ denoisers is comparable to the benchmark algorithms, with average PSNR values typically falling somewhere between those of the K-SVD and BM3D algorithms.  These results suggest that subspace-based denoising provides a reasonable venue for testing the capabilities of the \Nystrom\ covariance estimator.


\begin{table} 
	\begin{center}
		\caption{Average PSNR \up{(}in dB\up{)} of denoising algorithms for high-resolution test images}
		\label{tab:DenoisingHiRes}
		\vspace{1ex}
		\begin{tabular}{|c|r@{\,/\,}l|c|c|c|c|}
			\hline
				Image & \ \ $\sigma$ & PSNR & PCA & \Nystrom & K-SVD & BM3D \\
			\hline\hline 
			 	              & 10 & 28.13 & 37.81 & 38.80 & 34.84 & 43.76 \\
			 	\emph{flower} & 20 & 22.11 & 32.31 & 34.02 & 32.18 & 40.47 \\
			 	              & 50 & 14.15 & 24.50 & 26.69 & 28.73 & 36.06 \\
			\hline
			 	              & 10 & 28.13 & 27.99 & 27.26 & 31.39 & 35.39 \\
			 	\emph{leaves} & 20 & 22.11 & 27.00 & 26.15 & 27.56 & 31.61 \\
			 	              & 50 & 14.15 & 22.92 & 22.88 & 23.16 & 26.78 \\
			\hline 
			 	              & 10 & 28.13 & 37.18 & 37.54 & 32.58 & 42.51 \\
			 	\emph{hdr}    & 20 & 22.11 & 32.08 & 33.42 & 29.13 & 39.20 \\
			 	              & 50 & 14.15 & 24.44 & 26.55 & 25.42 & 34.76 \\
			\hline
			 	              & 10 & 28.13 & 33.47 & 33.47 & 30.51 & 34.19 \\
			 	\emph{deer}   & 20 & 22.11 & 30.58 & 31.46 & 26.67 & 33.33 \\
			 	              & 50 & 14.15 & 24.17 & 26.05 & 22.57 & 31.98 \\
			\hline
		\end{tabular}
	\end{center}
\end{table} 

 
In comparing the PCA and \Nystrom\ denoisers, we find that despite requiring significantly less computation, the \Nystrom\ approach actually performs slightly better in most cases.  One explanation for this behavior is that the shrinkage performed by the \Nystrom\ covariance estimator allows for improved subspace estimation.  

\begin{figure}
	\centering
	\includegraphics[width=0.5\textwidth]{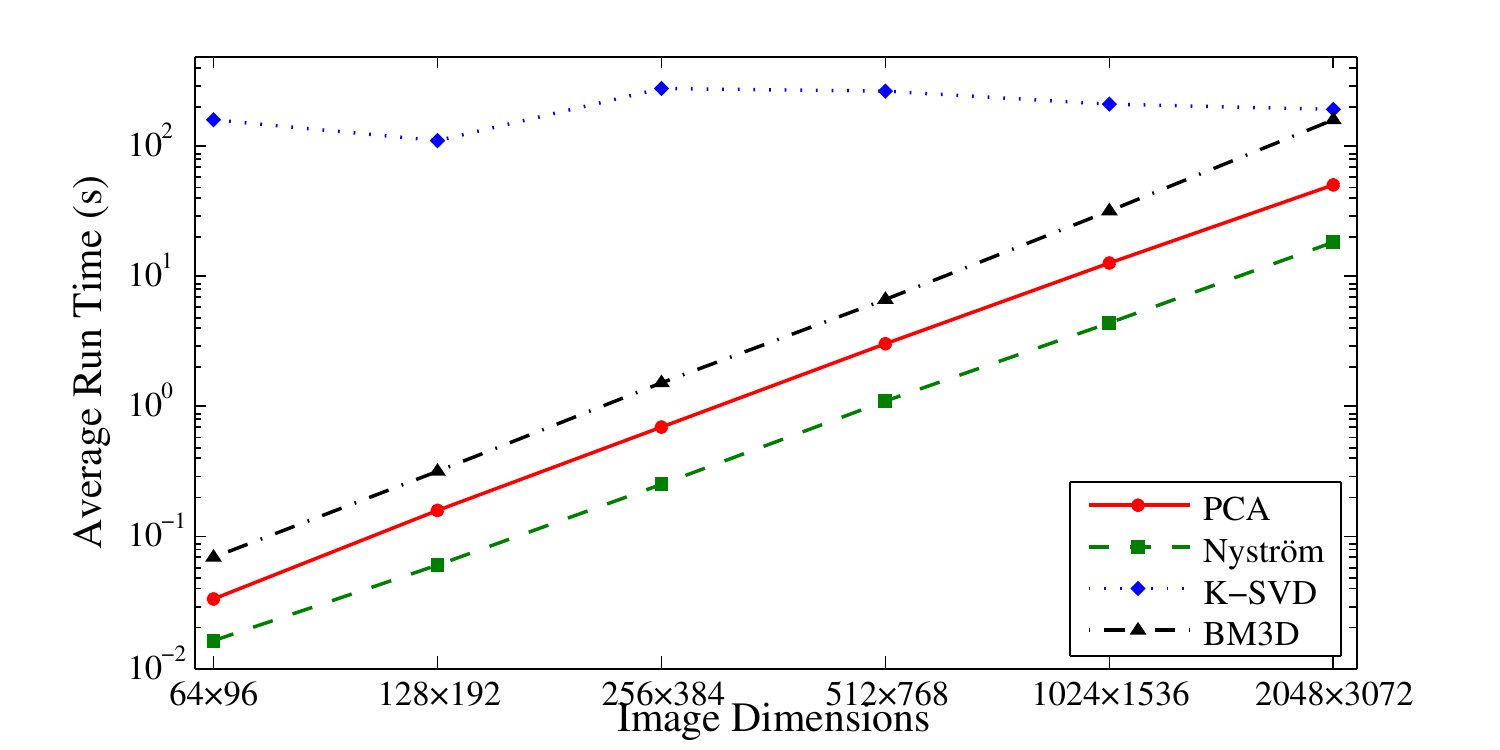}
	\caption{Average running times in seconds of various denoising algorithms when applied to different sizes of \emph{hdr} image.}
	\label{fig:ImageRunTimes}
\end{figure}

To better illustrate the computational differences between the algorithms, in \figref{fig:ImageRunTimes} we show the average run time of each method when applied to a noisy version of the \emph{hdr} image with $\sigma = 20$.  Results are shown for the original \mbox{$2048 \times 3072$} image, as well as for resampled versions with sizes of \mbox{$64 \times 96$}, \mbox{$128 \times 192$}, \mbox{$256 \times 384$}, \mbox{$512 \times 768$}, and \mbox{$1024 \times 1536$}.  We see that the costs of the PCA, \Nystrom\ and BM3D denoisers scale similarly with image size, with the \Nystrom\  denoiser performing about 2--3 times faster than the PCA denoiser and 3--9 times faster than BM3D.  Note that due to the complexity of its training step, the cost of K-SVD is high (though relatively constant) across all image sizes.


\begin{figure*}[t]
	\centering
	\includegraphics[width=0.3\textwidth, bb=0 0 256 256]{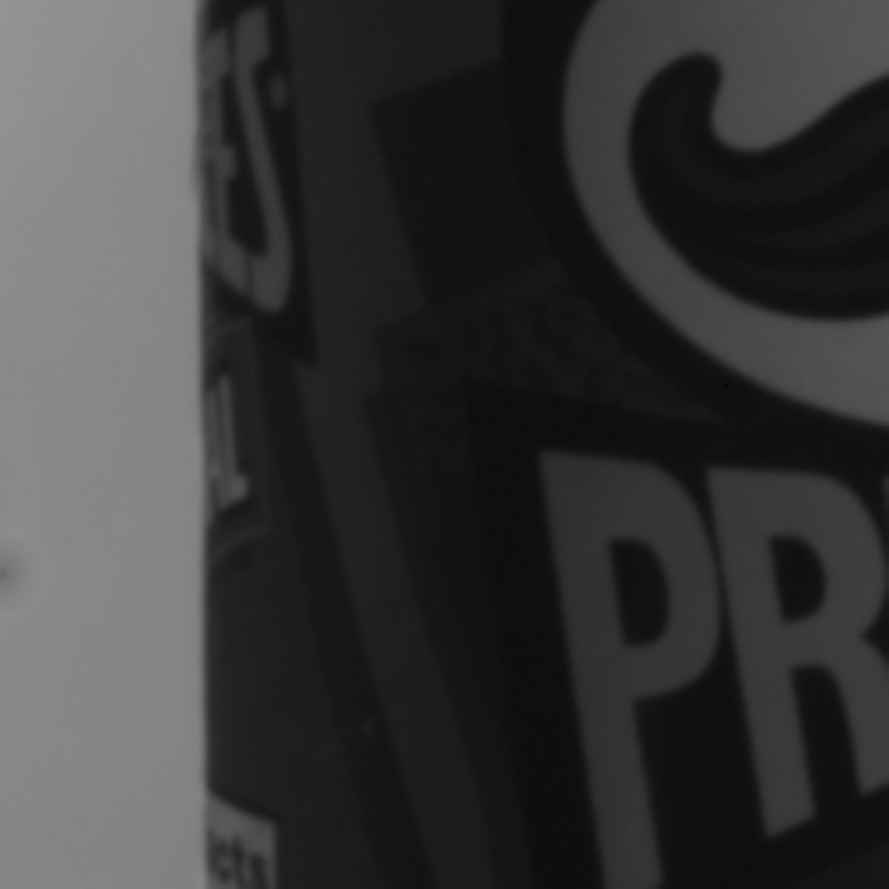}
	\hspace{0.05in}
	\includegraphics[width=0.3\textwidth, bb=0 0 256 256]{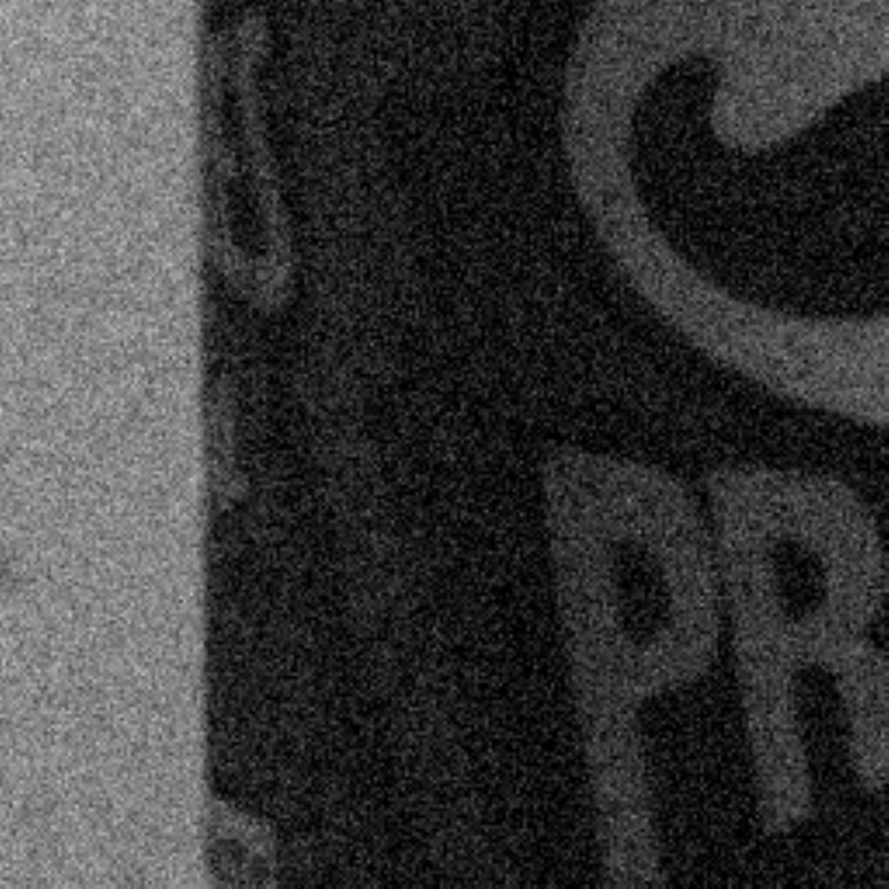}
	\hspace{0.05in}
	\includegraphics[width=0.3\textwidth, bb=0 0 256 256]{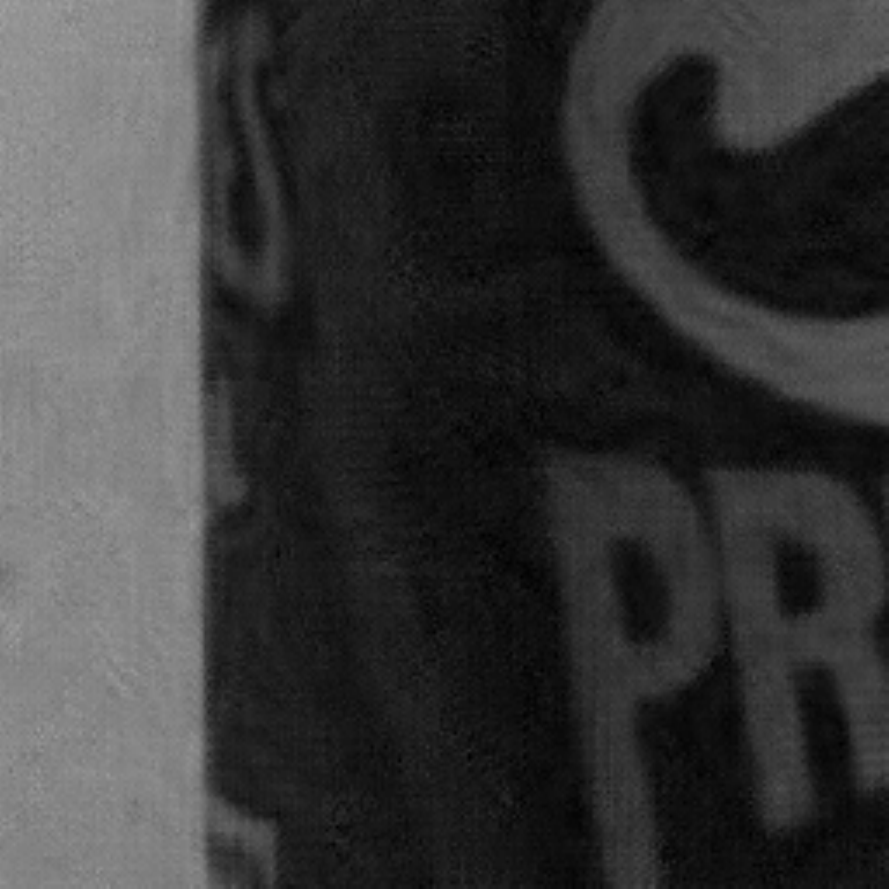}
	\\ \vspace{0.13in}
	\includegraphics[width=0.3\textwidth, bb=0 0 256 256]{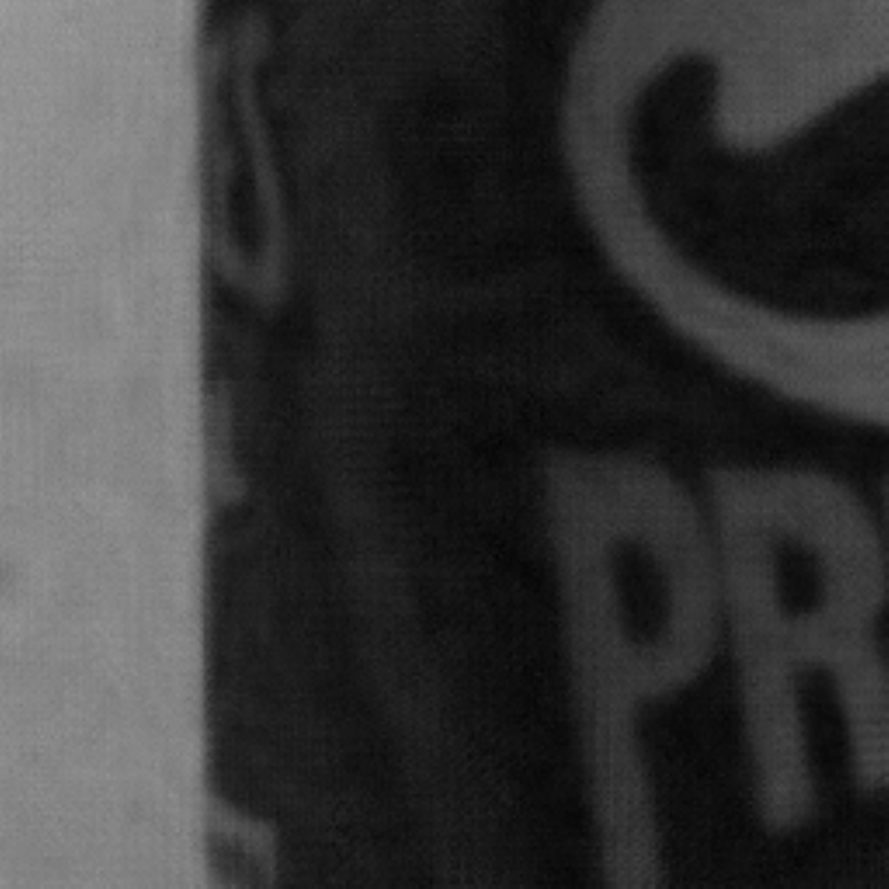}
	\hspace{0.05in}
	\includegraphics[width=0.3\textwidth, bb=0 0 256 256]{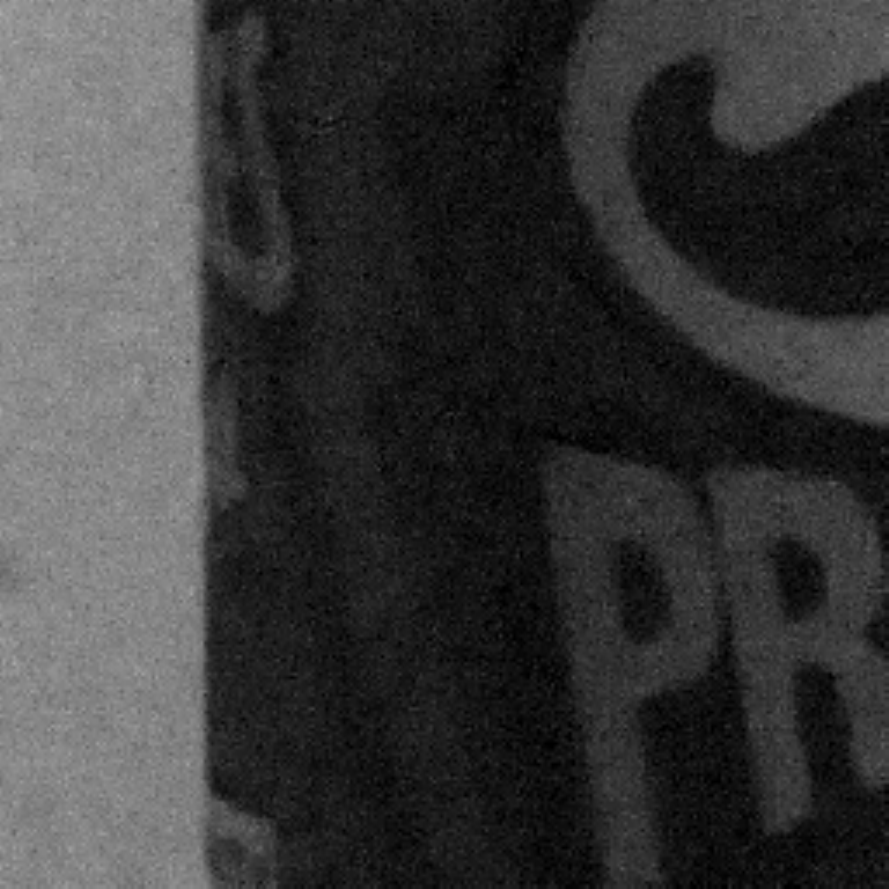}
	\hspace{0.05in}
	\includegraphics[width=0.3\textwidth, bb=0 0 256 256]{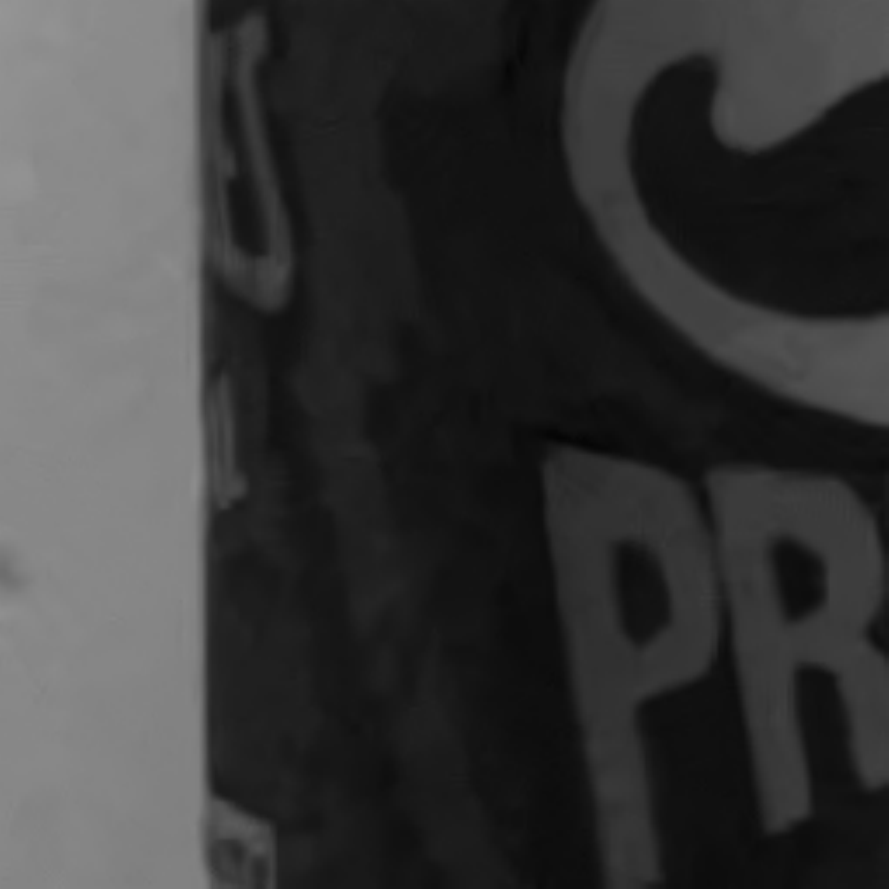}
	\caption{Close-up of denoising results for test image \textit{hdr}, given $\sigma = 20$.  Top left: original image.  Top right: noisy image (PSNR = 22.11 dB, SSIM = 0.947).  Middle left: PCA (PSNR~=~32.79~dB, SSIM~=~0.975).  Middle right: \Nystrom\ (PSNR = 34.35, SSIM~=~0.976).  Bottom left: K-SVD (PSNR~=~29.12~dB, SSIM~=~0.970).  Bottom right: BM3D (PSNR = 39.22 dB, SSIM = 0.986).}
	\label{fig:ImageComparison2}
\end{figure*}

In \figref{fig:ImageComparison2}, we show $256 \times 256$ close-ups of denoising results for the test image \textit{hdr}, given a noise level of $\sigma = 20$.  The \Nystrom-denoised image is visibly smoother and contains fewer artifacts than does the PCA-denoised image, as there is less mismatch between its estimated subspace and the ``true'' subspace represented by the clean image data.  This difference is reflected in the PSNR values for each method.  


\section{Summary} \label{sec:CovEstSummary}

In this article, we developed the \Nystrom\ approximation as a low-rank covariance estimator.  In addition to deriving expressions for its bias and mean squared error for the case of normally-distributed data, we showed that the \Nystrom\ covariance estimator shrinks the sample eigenvalues.  This shrinkage allows the estimator to achieve performance that is comparable to (or at times, better than) the sample covariance, particularly in cases where the number of samples is less than the dimension of the data.  Moreover, because of the computational advantages of the \Nystrom\ covariance estimator, its eigenvalues and eigenvectors can be computed for far less than the cost of spectral analysis of the sample covariance.

We illustrated the potential of the \Nystrom\ covariance estimator through its use in two example applications: array signal processing and image denoising.  In the first example, we adapted a projection-based beamforming algorithm to utilize the \Nystrom\ estimator, resulting in reduced computation with little degradation in our ability to recover the desired signal.  In the second example, we developed a simple PCA-based algorithm for image denoising, and then showed how the \Nystrom\ covariance estimator could be used to reduce computation while maintaining or even improving denoising performance.  


\bibliographystyle{IEEEtran}
\bibliography{ieeetsp2011}

\end{document}